\documentclass[a4paper,11pt]{article}
\pdfoutput=1 % if your are submitting a
\usepackage{jcappub} % for details on the use 
\usepackage{tensor}
\usepackage{enumitem}
\usepackage{titlesec}
\usepackage{makecell}
\usepackage{comment}
\usepackage{float}
\usepackage[table]{xcolor} % for row coloring
\usepackage{subcaption}
\usepackage{orcidlink}
\usepackage{multirow}
\usepackage{xcolor}
\usepackage{booktabs}
\usepackage{siunitx}
\usepackage{bm}
\usepackage{physics}
\usepackage{xspace}
\usepackage[normalem]{ulem}

\newcommand{\orcid}[1]{$\,$\href{https://orcid.org/#1}{\textcolor{orcidlogocol}{\faOrcid}}}

\newcommand{\Ms}{M_{\odot}}

\newcommand{\cW}{\mathcal{W}}

\newcommand{\cE}{\mathcal{E}}

\def\beq{\begin{equation}}
\def\eeq{\end{equation}}
\def\ber{\begin{eqnarray}}
\def\eer{\end{eqnarray}}
\def\benu{\begin{enumerate}}
\def\eenu{\end{enumerate}}

\title{\boldmath Impact of Anisotropy on Neutron Star Structure and Curvature}

\author[a]{A.~C.~Khunt \orcidlink{0000-0002-5615-4645},}
\author[b]{K.~Yavuz~Ek\c{s}i \orcidlink{0000-0001-5999-0553}}
\author[c]{ and P.~C.~Vinodkumar \orcidlink{0000-0003-2270-7691}}

\affiliation[a]{Department of Physics, Sardar Patel University, Vallabh Vidyanagar 388120, Gujarat, India}
\affiliation[b]{Istanbul Technical University, Faculty of Science and Letters, Physics Engineering Department, 34469 Istanbul, Turkey}
\affiliation[c]{P. D. Patel Institute of Applied Science, CHARUSAT, Changa 388421, Gujarat , India}

\emailAdd{ack.gravity@gmail.com }
\emailAdd{eksi@itu.edu.tr}
\emailAdd{p.c.vinodkumar@gmail.com}

\abstract{We investigate the impact of pressure anisotropy on the structural and geometric properties of neutron stars within general relativity, focusing primarily on the phenomenological Bowers–Liang (BL) model,  and comparing selected results with a quasi-local prescription. Using the SLy equation of state, we explore how anisotropic stresses modify global observables such as the mass–radius relation, moment of inertia, compactness, and tidal deformability over a broad range of anisotropy parameters. We find that moderate positive anisotropy can increase the maximum supported mass up to approximately $2.4\;M_\odot$ and enhance stellar compactness by up to $20\%$ relative to isotropic configurations, while remaining broadly consistent with current NICER and gravitational-wave constraints.

 To probe the internal gravitational field, we compute curvature invariants including the Ricci scalar, the Ricci tensor contraction, the Kretschmann scalar, and the Weyl scalar.  We show that curvature measures directly tied to the matter distribution exhibit a strong sensitivity to anisotropy, whereas the Weyl curvature remains comparatively insensitive, reflecting its role as a measure of the free gravitational field. Within the phenomenological BL framework, the maximum compactness increases with anisotropy and reaches values as high as $\mathcal{C}_{\max}\approx 0.25$-$0.38$  for $\lambda_{\rm BL}\in[-4,+4]$, although the physical realizability of such highly compact configurations depends sensitively on the underlying anisotropy mechanism. A comparison with the quasi-local model highlights the strong model dependence of anisotropic effects, underscoring both the potential significance and the limitations of phenomenological anisotropy prescriptions in modeling strong-field neutron-star interiors.
}

\arxivnumber{2512.24194}
\begin{document}
\maketitle
%\flushbottom
\section{Introduction}\label{sec:intro}
Neutron stars (NSs) offer unique laboratories for investigating matter under strong gravitational fields; however, their interior composition remains poorly understood.  In recent years, significant progress has been made in constraining fundamental parameters, such as mass, radius, and tidal deformability, using observational data. In this introduction, we present a brief summary of only a few of the most essential ones. 
The binary system PSR J0740+6620, which has a NS and a white dwarf, contains the most massive NS that has been found so far \cite{cromartie_2020}. With a mass of $2.08 \pm0.07 \Ms$, close to the predicted upper limit for NS masses  \cite{fonseca_2021}. 
Several studies have addressed the maximum mass of NSs utilizing data from gravitational wave (GW) observations (see recent reviews in Refs. \cite{li_2019,lim_2019,malik_2018}, as well as the references therein). Malik et al. \cite{malik_2018} have shown that the tidal deformability obtained from GW170817 could impose severe constraints on the nuclear matter equation of state (EoS) properties, such as the incompressibility slope and symmetry energy curvature. Lim and Holt \cite{lim_2019} conducted a thorough Bayesian investigation of the nuclear EoS, using a simple model at high densities and ruling out the possibility of phase transitions.  They investigated GW170817's tidal deformability and radius constraints using 300,000 NS EoSs generated from posterior probability distributions.  While the findings are mostly compatible with observational data, over 30\% of the models failed to fulfill the $2.0\;\Ms$ maximum mass criterion. This emphasizes the need for more robust microscopic constraints, notably from chiral effective field theory. The authors of Ref. \cite{li_2019} conclude that the maximum mass of NSs obeys the constraint $M_{\text{max}}\leq 2.4\;\Ms$, independent of the type of EoS.  This investigation suggests a possible threshold for separating NSs from black holes (BHs). In addition, precise estimates of NS radii are a crucial step toward further restricting the EoS. Additionally, X-ray bursts from accreting NSs in low-mass X-ray binaries (LMXBs) provide a means to constrain both the mass and radius of NSs.  As described in Ref. \cite{li_2019}, numerous studies in recent years have revealed similar radius limits. Nevertheless, accurately estimating NS radii remains challenging, principally because there are systematic inaccuracies in observational data \cite{ozel_2016}.

Improved Bayesian analyses employing NICER data for PSR J0437–4751 have since made it practicable to impose additional constraints on the radii and densities of NSs. These outcomes further strengthened the argument for a negative trace anomaly in massive NSs \cite{brandes_2025}.
LIGO and Virgo's GW detections of merging NSs provide measurements of their tidal deformability, leading to a useful instrument for probing a broad spectrum of NS masses as well as associated central densities \cite{abbott_2017,abbott_2018,abbott_2019,abbott_2020}. Two GW events from binary neutron star (BNS) mergers, GW170817 \cite{abbott_2017,abbott_2018} and GW190425 \cite{abbott_2020}, have been reported in recent years, granting significant constraints on the NS EoS.  Moreover, further investigations of the moment of inertia (MI) through pulsar timing offer an additional interesting trajectory for probing NS properties.

Despite the increasing number of NS observations, several aspects of their nature are still poorly understood. The uncertainty arises primarily from a lack of reliable understanding of matter's behavior at supranuclear densities and the effects of strong gravity. The study of matter at supranuclear densities is made possible by these extreme astrophysical laboratories, which connect strong-field gravity, particle physics, and nuclear physics \cite{lattimer_2004,potekhin_2010}. Although NSs are often represented as static spheres, rotation introduces oblateness, disrupting spherical symmetry and fundamentally altering their properties. Moreover, complex dynamics such as pressure anisotropy, which arise from unusual phenomena like superfluidity \cite{baym_1969,haskell_2018,heiselberg_2000,pines_1985}, phase transitions \cite{carter_1998,lombardo_2001}, ultra-strong magnetic fields \cite{yazadjiev_2012,ioka_2004,frieben_2012,kamiab_2015,folomeev_2015}, or crystalline cores \cite{nelmes_2012,canuto_1974}, make it challenging to understand their internal structure. Despite these developments, additional investigation is required to understand the interaction between gravity and extreme matter states in NS.

In the past few decades, theorists have devoted considerable effort to developing realistic EoS for dense matter in NSs. Yet there is still an important issue: most EoSs are based on quantum many-body theory in flat spacetime, whereas the structure of NSs depends on solving the Tolman–Oppenheimer–Volkoff (TOV) equations in curved spacetime. The TOV framework is based on general relativity (GR); however, the input EoS often neglects gravity's effects on matter at the microphysical scale. This raises another question of consistency: can flat-spacetime EoSs accurately portray matter under extreme gravity? If curvature influences the properties of dense matter, incorporating spacetime effects into EoS models, although exceedingly challenging, becomes crucial.

Although the local microphysics of dense matter can be accurately described using flat-spacetime nuclear EoS due to negligible metric variation over nuclear scales ($\sim 10^{-15}\;\rm m$)\cite{glendenning_2012}, recent research shows that quantum-field effects in curved spacetime can influence global thermodynamic quantities in ultra-compact or rapidly rotating NSs.  Gravitational time dilation, frame dragging, and spin-curvature coupling may alter the effective temperature, chemical potential, and polarization properties, resulting in small but potentially significant changes to the macroscopic EoS and stellar mass constraints \cite{hossain_2021a,hossain_2021b,hossain_2022,hossain_2023,li_2022}.

Ek\c{s}i et al.~\cite{eksi_2014} showed that compactness and curvature inside NSs are orders of magnitude greater than the values prevailing in the Solar System tests. Since the EoS is only slightly above the densities tested in terrestrial experiments, they argued that a mass-radius measurement is more likely to constrain gravity than the EoS. Using a relativistic mean-field (RMF) approach, one finds that dark matter significantly affects the interior curvature structure of Quarkyonic NSs, altering the central density, pressure, and compactness via a softened EoS.

The majority of current models assume isotropic pressure; however, pressure anisotropy can be caused by a number of factors, including phase transitions and strong interactions (see \cite{herrera_1997} for a review). Incorporating such anisotropy offers an improved account of the star's interior within GR.

Several recent investigations have used various theoretical frameworks to examine the influence of pressure anisotropy on NS properties.
Rahmansyah et al.~\cite{rahmansyah_2020} implemented GW170817 data to constrain anisotropy within the Bowers Liang \cite{bowers_1974}, Horvat et al. ~\cite{horvat_2010}, and Cosenza et al.~\cite{cosenza_1981} models, demonstrating that star radii are model-dependent. In modified gravity,~\cite{pretel_2022} investigated anisotropic stars in the  $f(R,T) = R + 2\beta T $ theory and observed that the coupling term enlarges stellar radii and allows for heavier masses.  Using scalar anisotropy models, ~\cite{das_2022} determined that positive anisotropy increases mass and radius while weakening the universal $I- \text{Love} -C$ relationship, with GW170817 data constraining the limits of the model. The Quasi-Local model ~\cite{mohanty_2024} showed significant implications on macroscopic and oscillation properties, whereas~\cite{das_2023} linked anisotropy to surface curvature (SC) using SC-$\Lambda$ and SC-$I$ relations. The study also found that anisotropy could increase the maximum mass of NSs by up to 15\%. This could potentially explain massive compact objects, for example, the secondary component of GW190814, and suggest universal relationships between binding energy, compactness, and anisotropy~\cite{becerra_2024}.

Recent work has emphasized the importance of general consistency criteria, such as energy conditions, causality, and stability, for anisotropic stellar models, highlighting that commonly used phenomenological prescriptions may satisfy these requirements only within restricted parameter ranges~\cite{becerra_2024,becerra_2025,pretel_2020,horvat_2010, mohanty_2024u}. In this spirit, the present study adopts anisotropy primarily as a controlled probe of strong-field effects, rather than as a micro-physically derived description of neutron-star matter

In this work, we investigate the influence of pressure anisotropy on the macroscopic, rotational, tidal, and geometric properties of neutron stars within the framework of general relativity. Adopting the SLy equation of state~\cite{douchin_2001}, we primarily employ the phenomenological Bowers–Liang anisotropy model to examine how deviations from isotropic pressure modify equilibrium structure, moment of inertia, tidal deformability, and spacetime curvature, and we compare selected results with those obtained using a quasi-local anisotropy prescription. 
The paper is organized as follows. In Sec.~\ref{sec:TF}, we present the theoretical framework governing anisotropic hydrostatic equilibrium, slow rotation, tidal perturbations, and curvature invariants. Numerical results for the mass–radius relation, moment of inertia, tidal properties, and curvature diagnostics are discussed in Sec.~\ref{sec:NAR}, together with an analysis of compactness bounds in anisotropic configurations. Our main findings and their implications, as well as the limitations of phenomenological anisotropy models, are summarized in Sec.~\ref{sec:Disc}.
Appendix~\ref{APD:QL} briefly presents some of the results based on the Quasi-Local (QL) anisotropic model for comparison. The Appendix contains further derivations for curvature and anisotropy at the star core, as well as other supplementary outcomes.

\textbf{\emph{Notations and Conventions}}: We use the metric signature $(-,+,+,+)$ and adopt geometric units with $G=c=\hbar=1$, while physical quantities are also quoted in cgs and natural units wherever appropriate.

\section{Theoretical framework} \label{sec:TF}
\subsection{Stable anisotropic compact star configurations}

\begin{figure}[H]
    \centering
    % First figure
    \begin{subfigure}[b]{0.45\textwidth}
        \centering
        \includegraphics[width=\textwidth]{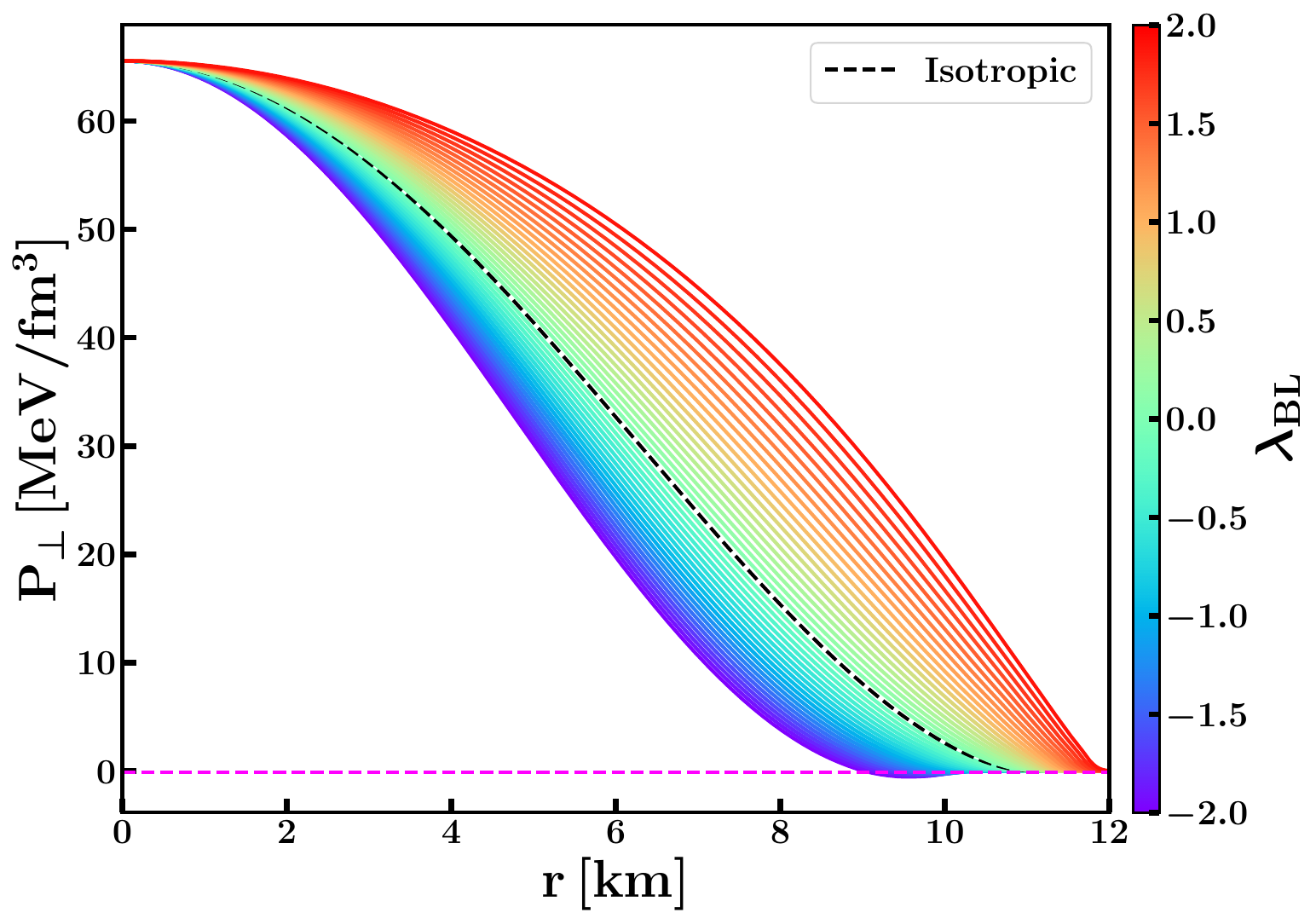}
        \caption{tangential pressure profile}
        \label{fig:tan_p}
    \end{subfigure}
    \hspace{0.01\textwidth} % small gap
    % Second figure
    \begin{subfigure}[b]{0.45\textwidth}
        \centering
        \includegraphics[width=\textwidth]{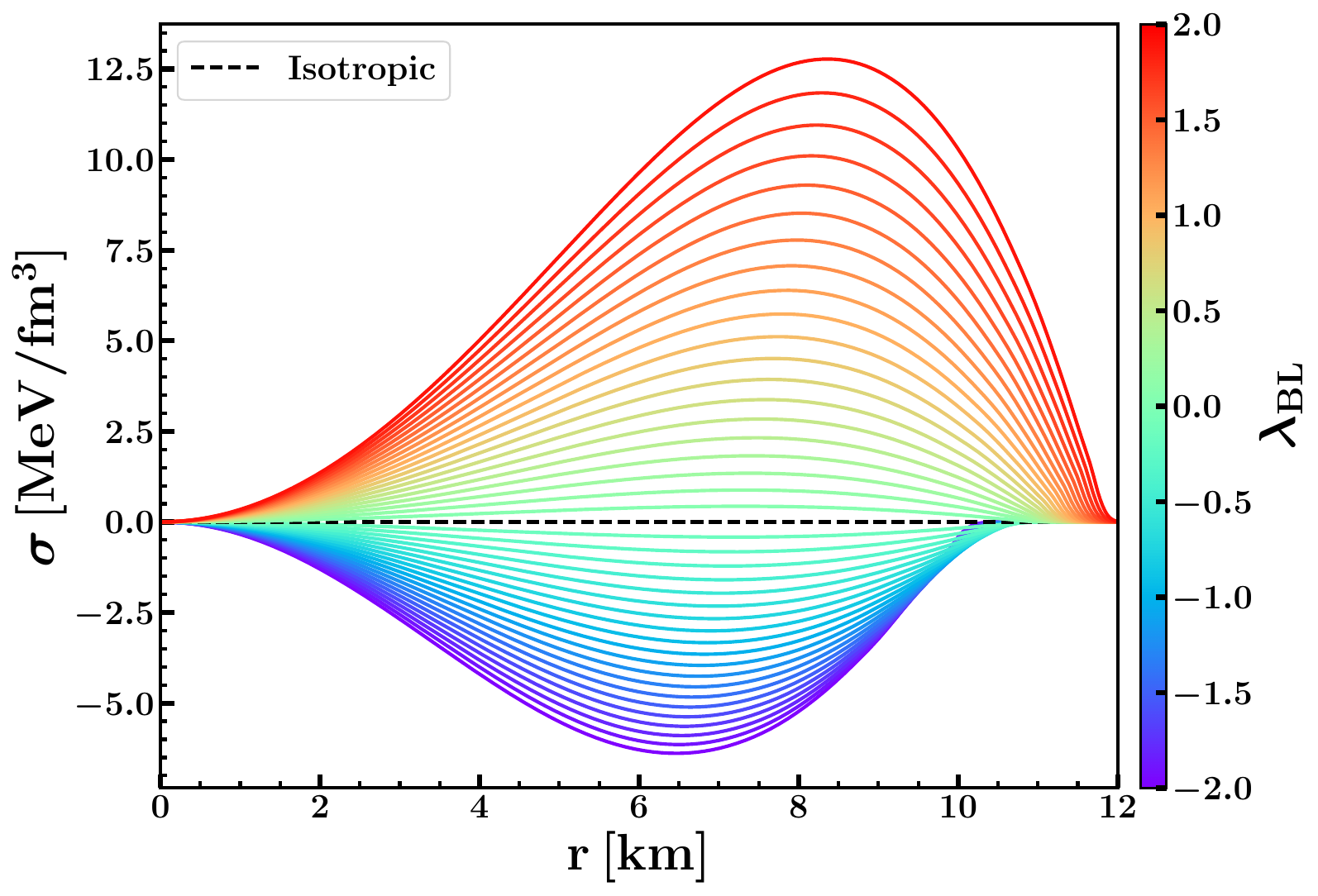}
        \caption{anisotropy profile}
        \label{fig:sigma}
    \end{subfigure}
    \caption{
\textbf{Panel (a):} Tangential pressure as a function of the radial coordinate for a star with different values of $\lambda_{\text{BL}}$. The black dashed line corresponds to the isotropic case. 
\textbf{Panel (b):} Anisotropy factor, $\sigma$, as a function of the radial coordinate for the SLy EoS.
}
\label{fig:tan_p_sigma}
\end{figure}

The stress-energy tensor for an isotropic star under ideal fluid conditions is defined as \cite{walecka_1974}
\begin{equation}\label{eq:tmunu}
    T^\mu{}_{\nu} = (\mathcal{E} + P)\, u^\mu u_\nu + P\, \delta^\mu{}_{\nu},
\end{equation}
where $\mathcal{E}$, $P$, and $u^\mu$ represent the energy density, pressure, and four-velocity of the ideal fluid, respectively.  
However, physical processes inside compact stars generally lead to anisotropy between the radial and tangential pressures.  
For a static, spherically symmetric anisotropic fluid distribution, the stress-energy tensor takes the diagonal form
\begin{equation}
    T^\mu{}_{\nu} = \mathrm{diag}(-\mathcal{E},\, P_r,\, P_{\perp},\, P_{\perp}) ,
\end{equation}
and can equivalently be written as
\begin{equation}\label{eq:anis_tmunu}
    T^\mu{}_{\nu}
    = (\mathcal{E}+P_{\perp}) u^\mu u_\nu 
      + (P_r-P_{\perp}) k^\mu k_\nu 
      + P_{\perp}\, \delta^\mu{}_{\nu},
\end{equation}
where $P_{\perp}$ and $P_r$ are the tangential and radial pressures, respectively.  
Here $k^\mu$ is the unit radial spacelike vector satisfying $k^\mu k_\mu = 1$ and $u^\mu k_\mu = 0$.
In Schwarzschild coordinates, the background metric is
\begin{align}
\mathrm{d}s^{2}
&= g^{(0)}_{\mu\nu}\, \mathrm{d}x^{\mu}\mathrm{d}x^{\nu} \nonumber\\
&= - e^{\nu(r)}\,\mathrm{d}t^{2}
   + e^{\lambda(r)}\,\mathrm{d}r^{2}
   + r^{2}\,\mathrm{d}\theta^{2}
   + r^{2}\sin^{2}\theta\,\mathrm{d}\varphi^{2}.
\end{align}
The anisotropic hydrostatic equilibrium equations that result from using this energy-momentum configuration and metric structure in GR are as follows:
\beq
 \frac{\mathrm{d}P_r}{\mathrm{d}r}=-\frac{\left({\cal{E}} + P_r \right)\left(m + 4\pi r^3 P_r \right)}{r\left(r -2m\right)} +\frac{2}{r} (P_{\bot} -P_r)\,,
  \label{tov1:eps}
\eeq
\beq\label{tov2:eps}
\frac{\mathrm{d}m}{\mathrm{d}r}=4\pi r^{2}{\cal{E}}\:,
\eeq
 and 
 \beq\label{eq:nu}
     \frac{\mathrm{d}\nu}{\mathrm{d}r} =\frac{m+4\pi r^3 P_r}{r(r-2m)}
 \eeq
where we define $\sigma=P_{\bot}-P_r$ as the anisotropy parameter. The $m$ stands for the mass within the radial coordinate $r$.  To close the system, there are two different EoSs for $P_r$ and $P_{\bot}$. We use a barotropic EoS for the radial pressure, $P_r = P_{r}({\cal{E}})$, and the BL model for the tangential pressure, $P_{\bot}$. 
The Runge-Kutta method is used to numerically solve the hydrostatic equilibrium equations for a given anisotropy. The starting point is the center of the star, where the boundary conditions are $r = 0$, $m = 0$, and $\mathcal{E} = \mathcal{E}_{c}$. 
 The integration continues outward until the pressure vanishes, which defines the surface of the star ($P(R)=0$), and at this point, $m(R) = M$. For the $P_{\bot}$, we adopt the BL model as presented in Ref. \cite{bowers_1974}
\beq\label{Anisotropy_eos}
    P_\bot = P_r + \frac{\lambda_{\rm BL}}{3} \frac{({\cal E}+3P_r)({\cal E} + P_r)r^2}{1-2m/r} \,,
\eeq
The parameter $\lambda_{\text{BL}}$ indicates the degree of anisotropy in the fluid. The following assumptions govern the model's construction: (i) the anisotropy must quadratically decrease to zero in the center, (ii) it needs to exhibit a nonlinear dependency on $P$, and (iii) gravitational effects are responsible for an element of the anisotropy. The parameter $\lambda_{\rm BL}$ controls the fluid's degree of anisotropy. To maintain spherical symmetry and ensure the physical consistency of anisotropic NS models, all standard acceptability conditions must be satisfied. We have explicitly verified that our configurations fulfill these requirements, including regularity, energy conditions, and causality throughout the stellar interior. The corresponding results for the SLy parameter set are presented in Fig.~\ref{fig:tan_p_sigma}.

The radial variation of the tangential pressure for a given central density is shown in Fig.~\ref{fig:tan_p}.  It is evident that $P_\perp$ increases with increasing $\lambda_{\rm BL}$, indicating the potential to sustain more massive NS configurations, whereas decreasing $\lambda_{\rm BL}$ leads to lower-mass comparisons. The condition is satisfied in the star core where $P_\perp$ and $P_r$  coincide.  Positive values of $\lambda_{\rm BL}$ result in a positive $P_\perp$ toward the surface, while negative values have an opposite effect. The nonphysical solutions are mostly found near the surface due to this negative value.  Fig.~\ref{fig:sigma} for maximum mass NS provides a clearer view of the anisotropy parameter's magnitude.  When comparing the maximal star to the canonical star, the negative magnitude increases. Since $(1 - e^{-\lambda} = 2m/r) \sim r^2$ as $r \rightarrow 0$, the BL model with gravitational anisotropy has the important characteristic role of providing isotropy of the fluid at the star center.  Moreover, this model considers only relativistic cases, in which extreme densities produce gravitational anisotropy \cite{bowers_1974}. 
%{We explored and illustrated in Fig.~\ref{fig:APD_Sigma_cont} (refer to Appendix) the contour representation of the anisotropy factor ($\sigma$), demonstrating its variation and evolution in relation to the radial coordinate and the anisotropy parameter $\lambda_{\rm BL}$ of the BL model.}

\begin{figure}[H]
    \centering
    \includegraphics[width=0.45\linewidth]{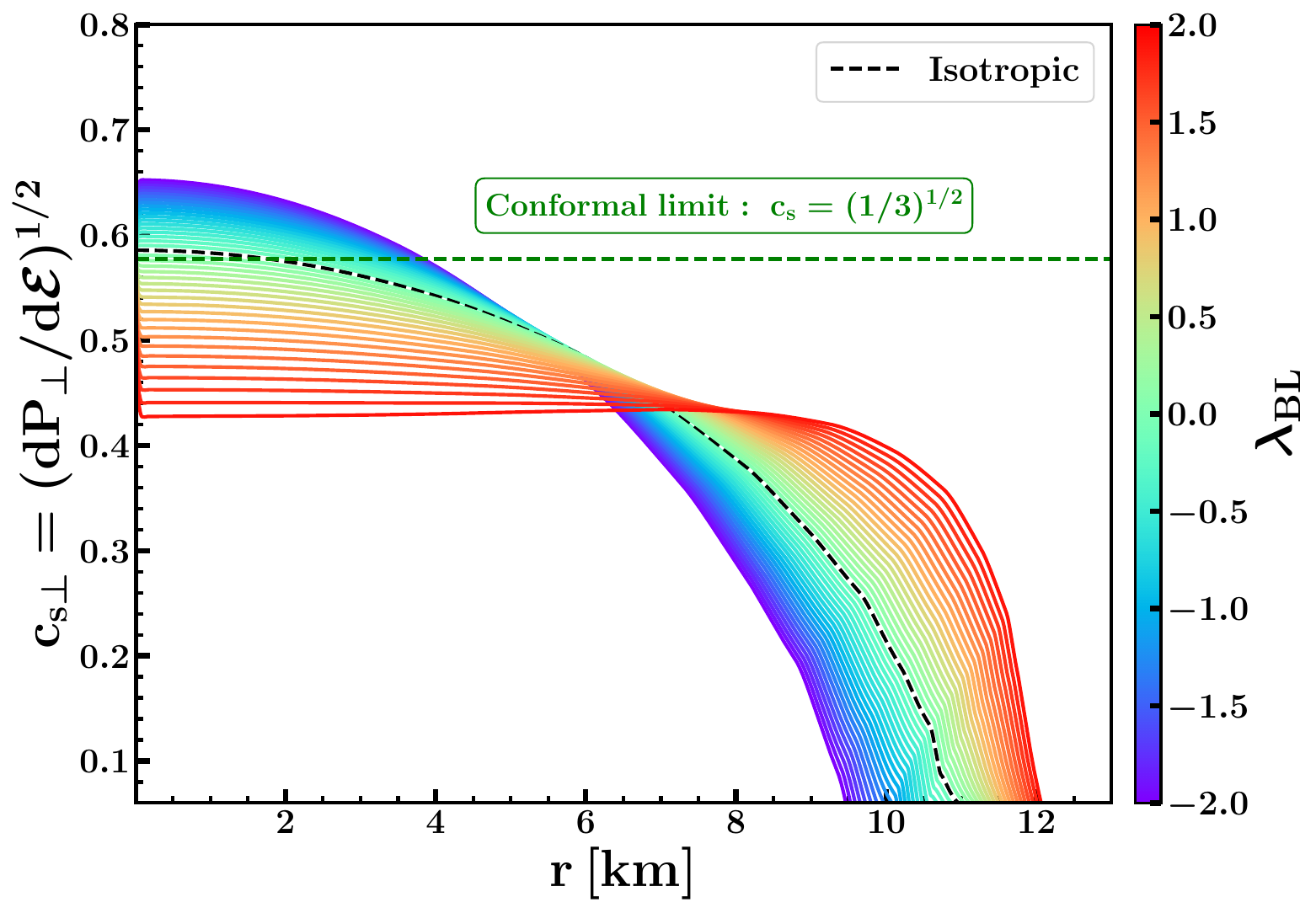}
    \caption{The radial profile of the tangential sound speed ($c_{s\perp}$) with the SLy EoS for various $\lambda_{\rm BL}$ values at fixed central density.  The conformal limit is shown by the dashed green line.}
    \label{fig:c_st}
\end{figure}

Fig.~\ref{fig:c_st} depicts the radial profile of the tangential sound speed, which is defined as $c_{s\perp} = (\partial P_\perp/ \partial \mathcal{E})^{1/2}$. For various values of the anisotropy parameter $\lambda_{\rm BL} \in [-2.0, +2.0]$, the radial change in the $c_{s\perp}$ is demonstrated. The conformal limit $(c_s = c/\sqrt{3} \approx 0.577c)$ is provided for comparison. In the isotropic case and with positive anisotropy ($\lambda_{\rm BL} > 0$), the sound velocity remains below the conformal limit in the stellar interior. The possibility of an ultra-stiff matter distribution in the core can be seen by the tangential sound velocity exceeding the conformal limit across the inner $\sim 4\,\text{km}$ region from the stellar center for negative anisotropy ($-2.0 \leq \lambda_{\rm BL} \leq -0.4$). The $c_s $ decreases while approaching the conformal limit beyond this radius. As well, the $c_{s\perp}$ is found to be consistent with the causality condition throughout the stellar interior in the range $-2 < \lambda_{\rm BL} < 2$ for the maximum mass configuration given by the SLy EoS.

\subsection{Mathematical representation of spacetime curvatures}
In this part, we provide the mathematical expressions for the scalar curvature invariants, including the Ricci scalar for a spherically symmetric NS, the Kretschmann scalar, and the full contractions of the Weyl and Ricci tensors. The expressions for these curvature quantities are given by \cite{eksi_2014,das_2021}.

\subsection*{\centering\mdseries\itshape 1. Kretschmann Scalar ($\mathcal{K}$)}
The Kretschmann scalar (full contraction of the Riemann tensor) is given by:
\beq
\mathcal{K}(r) \equiv \sqrt{R_{\mu\nu\rho\sigma} R^{\mu\nu\rho\sigma}},
\eeq
where the Riemann curvature tensor is shown by $R_{\mu\nu\rho\sigma}$.  The Kretschmann scalar for the anisotropic configuration could be obtained by calculating the contraction within the context of the modified TOV equations, using the Christoffel symbols and associated Riemann tensor components:
\beq\label{eqK}
\mathcal{K}(r) = \Big[(8\pi)^2 \left(3 \mathcal{E}^2 + 3 P^2 + 2 \mathcal{E} P\right) - \frac{128 \pi \mathcal{E} m(r)}{r^3}
+ \frac{48 m^2(r)}{r^6}\,\Big]^{1/2}\,,
\eeq
In this case, the mass encompassed by a radius $r$ is denoted by $m(r)$.  At $r > R$, the star's outer region, where $P=0$, $\mathcal{E}=0$, and $m(r)=M$, the expression simplifies to
\beq
\mathcal{K}(r) = \frac{4\sqrt{3} M}{r^3}.
\eeq
Similarly, at the star surface ($r = R$), we obtain
\beq
\label{Ksurface}
\mathcal{K}(R) = \frac{4 \sqrt{3} M}{R^3}.
\eeq

\subsection*{\centering\mdseries\itshape 2. Weyl Scalar  ($\mathcal{W}$)}

The Weyl tensor contraction is defined as:
\begin{equation}\label{Weyl}
\mathcal{W}(r) \equiv \sqrt{C_{\mu\nu\rho\sigma} C^{\mu\nu\rho\sigma}},
\end{equation}
where $C_{\mu\nu\rho\sigma}$ denotes the Weyl tensor. It represents the traceless part of the Riemann curvature tensor and is related to the latter through the following decomposition:
\ber
&  & C_{\mu\nu\rho\sigma} = R_{\mu\nu\rho\sigma}  +\frac{R}{6}\left(g_{\mu\rho}g_{\nu\sigma} - g_{\mu\sigma}g_{\nu\rho}  \right) \nonumber \\ 
& &  +\frac{1}{2} \left(R_{\mu\sigma}g_{\nu\rho} -R_{\mu\rho}g_{\nu\sigma} + R_{\nu\rho}g_{\mu\sigma} -R_{\nu\sigma}g_{\mu\rho} \right)\,. 
\eer
In addition to having the same symmetry features as the Riemann tensor, the Weyl tensor $C_{\mu\nu\rho\sigma}$ must also be trace-free, which means that any contraction with the metric over a set of indices vanishes in the exact same manner.

The following equation is obtained by contracting and evaluating the Weyl tensor's components for the spherically symmetric case:
\begin{align}
C_{\mu\nu\rho\sigma} C^{\mu\nu\rho\sigma}
&= \frac{4}{3} \left( \frac{6 m(r)}{r^3} - \kappa \mathcal{E}\right)^2 \\
&= \frac{4}{3} \left( \frac{6 m(r)}{r^3} - 8\pi \mathcal{E} \right)^2.
\end{align}
where $\kappa = 8\pi G$ (with $c=1$), and in geometrized units $\kappa = 8\pi$. Therefore, the scalar that comes from the Weyl tensor's full contraction is
\beq\label{eqW}
\mathcal{W}(r) = \left[ \frac{4}{3} \left( \frac{6 m(r)}{r^3} - 8\pi \mathcal{E}(r) \right)^2 \right]^{1/2}\,.
\eeq
In the exterior region of the star ($r > R$), the equation reduces to
\beq
\mathcal{W}(r) = \frac{4 \sqrt{3} M}{r^3}.
\eeq
At the star surface ($r = R$), the expression becomes
\beq\label{Wsurface}
\mathcal{W}(R) = \frac{4 \sqrt{3} M}{R^3}\,,
\eeq
Note that at the star surface $(r=R)$, the boundary conditions 
$P(R)=0$, $\mathcal{E}(R)=0$, and $m(R)=M$ hold. Under these conditions, 
the same form is obtained by reducing  Eqns.~\eqref{eqK} and \eqref{eqW}.
\beq
\mathcal{W}(R)=\mathcal{K}(R)=\frac{4\sqrt{3}\,M}{R^{3}}.
\eeq
\subsection*{\centering\mdseries\itshape 3. Ricci Scalar ($\mathcal{R}$)}
The Ricci scalar $\mathcal{R}$ is calculated by contracting the Ricci tensor $R_{\mu\nu}$, which is itself obtained from the Riemann curvature tensor.  It tells us how curved space-time is because of the distribution of mass and energy, and it is described as
\beq
\mathcal{R} = g^{\mu\nu} R_{\mu\nu}.
\eeq
The Ricci scalar is a scalar measure of curvature that shows how volumes vary in curved spacetime.  It may be immediately derived from the Einstein field equations by taking the trace, resulting in
\beq
\mathcal{R}(r)=\kappa\left(\mathcal{E}-3P   \right)\,,
\label{eq:Ricci_scalar}
\eeq
the full contraction of the Ricci tensor
\beq\label{ricci_tensor}
{\cal J} \equiv \sqrt{{\cal R}_{\mu \nu} {\cal R}^{\mu \nu} }= \kappa \left( \mathcal{E} + 3P\right).
\eeq
In contrast to the Kretschmann and Weyl scalars, the Ricci scalar and the Ricci tensor vanish at the stellar boundary and throughout the exterior vacuum, with their contributions confined entirely to the star’s interior. It is worth noting that, unlike ${\cal K}$, ${\cal W}$, and ${\cal J}$, 
the Ricci scalar ${\cal R}$ is not positive definite. From Eq.~\ref{eq:Ricci_scalar}, it follows that ${\cal R}$ may become negative in regions where $P > \mathcal{E}/3$. Such behavior reflects the dominance of relativistic pressure in the trace of the energy–momentum tensor and does not indicate any geometrical pathology. Rather, it signals the transition to a strongly relativistic matter regime in the stellar interior.

The above expressions also provide a useful consistency check. 
Since the Riemann tensor can be decomposed into Weyl and Ricci parts, the associated invariants obey the relation
\begin{equation}
{\cal K}^2 = {\cal W}^2 + 2{\cal J}^2 - \frac{{\cal R}^2}{3}.
\end{equation}
This relation shows that the curvature invariants are not independent. While ${\cal K}$ encodes the total curvature strength, ${\cal W}$ isolates the traceless (tidal) contribution, and ${\cal J}$ together with ${\cal R}$ describe the trace part sourced directly by matter. Therefore, each invariant provides complementary geometric 
information about the spacetime structure inside the star~\cite{cherubini_2002,eksi_2014}.

\subsection*{\centering\mdseries\itshape 4. Surface Curvature (SC)}
The total mass, $m \rightarrow M$, is obtained from the mass function at the stellar surface as $r \rightarrow R$. Since the Ricci scalar ${\cal R}$ and the invariant ${\cal J}$ depend on the energy density ${\cal E}(r)$ and pressure $P(r)$, both of which vanish in the vacuum region, they are zero outside the star. Nevertheless, the exterior spacetime remains curved, as indicated by the nonzero Riemann tensor component $\tensor{{\cal R}}{^1_{010}} = -2M/R^3$, showing that the Riemann tensor provides a more appropriate measure of curvature for compact stars.
The square root of its full contraction defines the Kretschmann scalar. In the vacuum exterior, both ${\cal K}$ and ${\cal W}$ reduce to $4\sqrt{3}M/R^3$, making them suitable curvature diagnostics. We define the surface curvature (SC) as ${\cal K}(R)/{\cal K}_\odot$, where ${\cal K}_\odot \equiv 4\sqrt{3}M_\odot/R_\odot^3 = 3.06 \times 10^{-27}\,\mathrm{cm}^{-2}$. For NSs, $\mathrm{SC} \approx 10^{14}$, indicating surface curvatures about fourteen orders of magnitude stronger than that of the Sun (see Fig.~\ref{fig:IC_SC}).

\subsection{Slow rotating neutron stars and moment of inertia}\label{sec:sub_MI}
This section presents a short overview of the main features of slowly rotating NSs with anisotropic pressure.

To determine the equilibrium configuration of a rotating star, one must solve the Einstein equations $G^{\mu}_{\nu} = 8\pi  T^{\mu}_{\nu}$. The selected line element provides the explicit form of the components of $G^{\mu}_{\nu}$ as
\begin{eqnarray}\label{rotat}
      \mathrm{d}s^2 = &-e^{2\nu(r)} \mathrm{d}t^2 + e^{2\lambda(r)} \mathrm{d}r^2 
       + r^2 (\mathrm{d}\theta^2 + \sin^2\theta \, \mathrm{d}\phi^2) 
       - 2\omega(r) \, r^2 \sin^2\theta \,\mathrm{ d}t \, \mathrm{d}\phi  
\end{eqnarray}
 with the frame-dragging effect taken into account, $\omega(r) \equiv (d\phi/dt)_{\rm ZAMO}$ is the Lense–Thirring angular velocity as determined by a zero-angular-momentum observer (ZAMO). The slow-rotation approximation necessitates $\Omega/\Omega_k \ll 1$ for a uniformly rotating NS with angular velocity $\Omega$ and Keplerian limit $\Omega_k$.
Therefore, the line element in Eq.(\ref{rotat}) is valid only up to first order in $\Omega$, with the star remaining spherical since centrifugal deformation appears at $\mathcal{O}(\Omega^2)$ \cite{idrisy_2015}.  The stress–energy tensor for anisotropic matter is expressed in Eq.~\eqref{eq:anis_tmunu}. The mass of the NS is simply $m(R)=M$. For the differential equation in Eq.~(\ref{eq:nu}), the provided boundary condition is $\nu$ at $R$, i.e., $e^{\nu(R)}= (1-\frac{2M}{R})$, in contrast with the ones in Eq.~\eqref{tov2:eps}, where the mass and pressure values at $r_c$ ($r=0$) are given. This differential equation is solved by assigning a trial value of $\nu$ at $r_c$ and integrating repeatedly until the condition for $\nu$ at $R$ is satisfied. Additionally, the $t\phi$-component of Einstein's field equation for the metric in Eq.~\eqref{rotat} has a second-order ordinary differential equation for $\bar{\omega}$ as
\begin{equation}
\frac{1}{r^{4}} \frac{\mathrm{d}}{\mathrm{d}r} \left( r^{4} J \frac{\mathrm{d}\bar{\omega}}{\mathrm{d}r} \right) 
+ \frac{4}{r} \frac{\mathrm{d}J}{dr} \left(1+ \frac{\sigma}{\mathcal{E}+P}\right)\, \bar{\omega} = 0 , 
\end{equation}
where
\begin{equation}
J\equiv e^{-\nu} \left (1 - \frac{2M}{R}\right)^{1/2} \, ,
\end{equation}
 where $\bar{\omega}$ is the frame dragging angular frequency, $\bar{\omega}(r) = \Omega - \omega(r)$.
The moment of inertia of a slowly rotating anisotropic NS was estimated in Ref. [39] as
\begin{align}
    I &= \frac{8\pi}{3}\int_0^R \frac{r^5J\Tilde{\omega}}{r- 2M}({\cal{E}}+P)\left[1+\frac{\sigma}{{\cal{E}}+P}\right] \, \mathrm{d}r,
    \label{eq:MI}
\end{align}
where $\Tilde{\omega}=\Bar{\omega}/\Omega$ and $\sigma = P_\bot - P_r = \frac{\lambda_{\rm BL}}{3} \frac{({\cal E}+3P_r)({\cal E} + P_r)r^2}{1-2m/r}$.
Accordingly, Eq.~(\ref{eq:MI}) can be reformulated with the aid of Eq.~(\ref{Anisotropy_eos}) as
\begin{align}\label{eq:MOI}
    I &= \frac{8\pi}{3}\int_0^R \frac{r^5J\Tilde{\omega}}{r- 2M}({\cal{E}}+P)\left[1+\frac{\frac{\lambda_{\rm BL}}{3} \frac{({\cal E}+3P)({\cal E} + P)r^2}{1-2m/r}}{{\cal{E}}+P}\right] \, \mathrm{d}r,
\end{align}
In Sec.~\ref{sec:MOI}, we present the computed MOI values for anisotropic NSs using the SLy EOS.

\subsection{Tidal deformability and Love numbers}
A quadrupole moment $\epsilon_{ij}$ is induced in a spherically symmetric star when it becomes subjected to a static external quadrupolar tidal field $Q_{ij}$.  First-order linear relationships exist between the two as \cite{hinderer_2008}
\begin{align}
    Q_{ij} = - \bm{\lambda} \epsilon_{ij}~.
\end{align}
The degree of star deformation caused by an external tidal force is described by the parameter $\bm{\lambda}$, commonly referred to as the (dimensionful) tidal deformability. The background metric $g_{\mu\nu}^{(0)}$ is perturbed by $h_{\mu\nu}$ when this field is present, allowing the entire metric to be written as
\begin{align}
    g_{\mu\nu} = g_{\mu\nu}^{(0)} + h_{\mu\nu}~.
\end{align}
The perturbation can also be separated into polar and axial components,
\begin{align}
     h_{\mu\nu} =  h_{\mu\nu}^{\rm polar} +  h_{\mu\nu}^{\rm axial}~.
\end{align}
The polar element of the linearized metric perturbation, \(h_{\mu\nu}^{\text{polar}}\), can be expressed as \cite{regge_1957,thorne_1967}
\begin{align} \label{polar_metric_perturbation}
    h_{\mu\nu}^{\text{polar}} =~& \text{diag} [e^{\nu (r)} H_0(r), e^{\lambda (r)} H_2(r), r^2 K(r), \\\nonumber& r^2 \sin^2\theta K(r)]\times Y_{lm}(\theta,\phi)~,
\end{align}
In this case, $Y_{lm}(\theta,\phi)$ represents for the spherical harmonics. The nonzero components of the stress-energy tensor are
\begin{align*}
    \delta T_0^0 &= -\delta \mathcal{E} = -\left(\dv{P}{\mathcal{E}}\right)^{-1} \delta P\,, \\
    \delta T_i^i &= \delta P~, 
\end{align*}
where $\delta P$ and $\delta \mathcal{E}$ represent the perturbations in pressure and energy-density, respectively. By substituting these expressions together with Eq.~\eqref{polar_metric_perturbation} into the linearized Einstein equations, it is determined that $H_2(r) = H_0(r) \equiv H(r)$~\cite{hinderer_2008}. The relationship between the dimensionless Love  number $(k_2)$ and the tidal deformability is $\bm\lambda = \tfrac{2}{3} k_2 R^5$, where $R$ corresponds to the star radius.
To determine $k_2$, we use linear perturbations within the Thorne–Campolattaro metric \cite{thorne_1967}. The following second-order differential equation for an anisotropic star is obtained by solving the Einstein equations \cite{biswas_2019}.
\begin{align}
H^{''} &+ H^{'} \bigg[\frac{2}{r} + e^{\lambda} \left(\frac{2m(r)}{r^2} + 4 \pi r (P - {\cal E})\right)\bigg] 
\nonumber \\
&
+ H \left[4\pi e^{\lambda} \left(4 {\cal E} + 8P + \frac{{\cal E} + P}{dP_\bot/d{\cal E}}(1+c_s^2)\right) -\frac{6 e^{\lambda}}{r^2} - {\nu^\prime}^2\right] 
\nonumber \\
&
= 0\,.
\end{align}
According to the energy density at constant $\lambda_{\rm BL}$, $dP_\bot/d\mathcal{E}$ represents the variation of $P_\bot$ (see Eq.~\eqref{Anisotropy_eos}).
In order to determine the tidal Love number \cite{damour_2009,hinderer_2008}, the internal and exterior solutions of the perturbation variable $H$ are matched at the surface of the star. We get its value using the function $y_2$ and the compactness parameter $\mathcal{C}$, which has been defined as ~\cite{hinderer_2008, Hinderer_2009}
\begin{align}
    k_{2}^{\text{polar}}(\mathcal{C},y_2) &= \, \frac{8}{5} \mathcal{C}^5 (1-2\mathcal{C})^2 \big[ 2(y_2-1)\mathcal{C} - y_2 + 2 \big]
    \nonumber \\ &
    \times \Big\{ 2\mathcal{C} \big[ 4(y_2+1)\mathcal{C}^4 + 2(3y_2-2)\mathcal{C}^3 - 2(11y_2-13)\mathcal{C}^2 
    \nonumber \\ &
    + 3(5y_2-8)\mathcal{C} - 3(y_2-2) \big]+ 3(1-2\mathcal{C})^2 
    \nonumber \\ &
    \times \big[ 2(y_2-1)\mathcal{C}-y_2+2 \big] \log(1-2\mathcal{C}) \Big\}^{-1} \, , 
    \label{eq:k2}
\end{align}
Here, $\mathcal{C} = M/R$ represents dimensionless compactness, and $y_2 = y(R)$. Where $y_2$ is determined by the surface value of $H$ and its derivative
\begin{equation}
    y_2 = \frac{rH^{'}}{H}\Big|_R.
\end{equation}
Using the equation above, one can obtain the dimensionless tidal deformability from the following expression,
\begin{equation} \label{Lambda}
    \Lambda = \frac{2}{3}k_2^{\text{polar}} \left(\frac{M}{R}\right)^{-5}.
\end{equation}
Finally, $\Lambda$ can be obtained from $k_2$, providing an important value to compare with the findings inferred  from GW170817 and GW190425.

\section{Numerical analysis and results}\label{sec:NAR}
\subsection{Mass-radius relations}\label{sec:sub_mr}
\begin{figure}[H]
    \centering
    % First figure
    \begin{subfigure}[b]{0.45\textwidth}
        \centering
        \includegraphics[width=\textwidth]{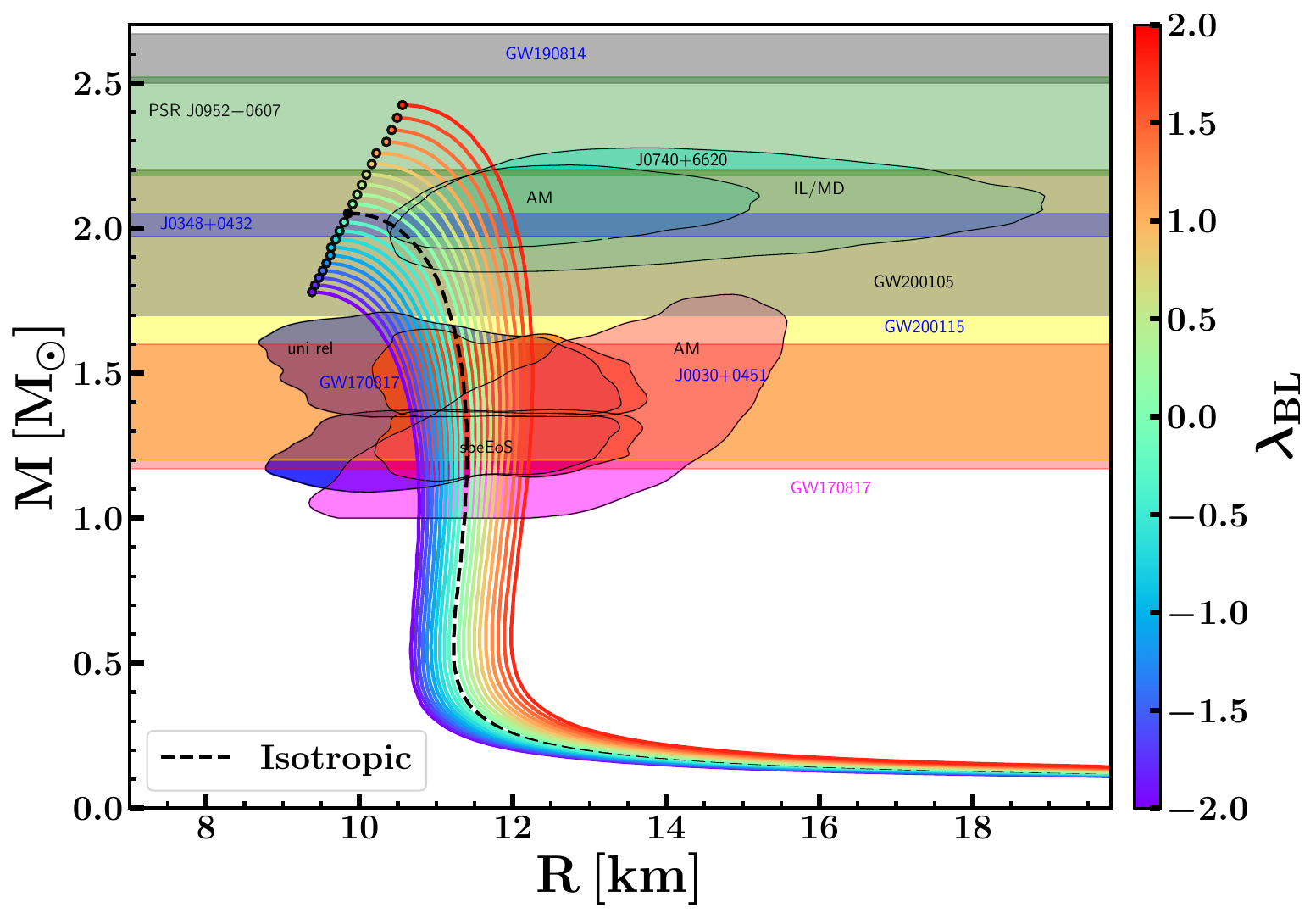}
        \caption{mass-radius relation}
        \label{fig:MR}
    \end{subfigure}
    \hspace{0.01\textwidth} % small gap
    % Second figure
    \begin{subfigure}[b]{0.45\textwidth}
        \centering
        \includegraphics[width=\textwidth]{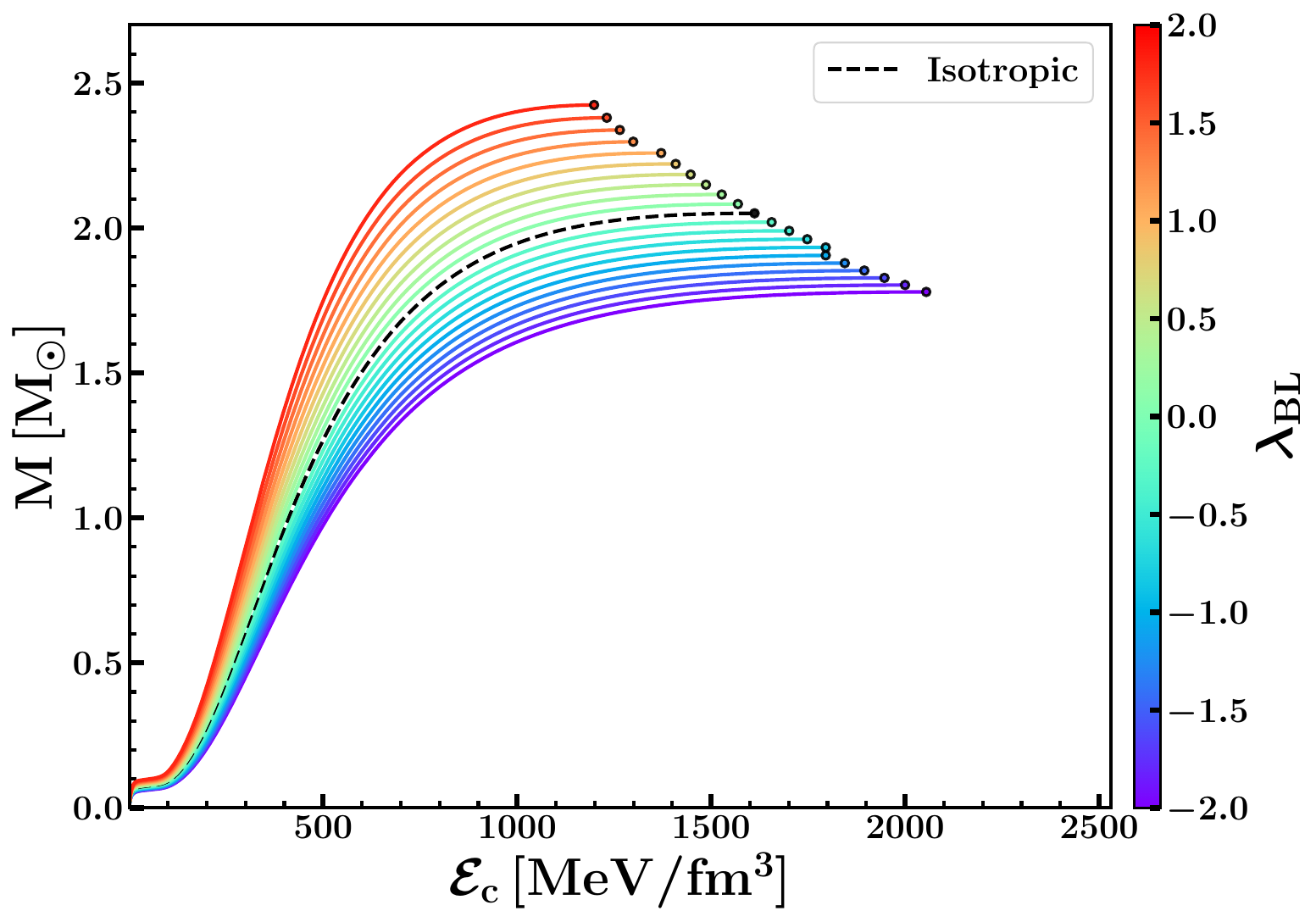}
        \caption{mass vs. central density}
        \label{fig:MR_rho}
    \end{subfigure}
    \caption{\textbf{Panel (a):} Impact of anisotropy on the mass–radius relation of NSs, illustrated for the BL model. The plot includes constraints from recent observations, namely the NICER measurements of PSR J0030+0451 \cite{riley_2019nicer,miller_2019psr}, PSR J0952-0607 \cite{Romani_2022} and PSR J0740+6620 \cite{miller_2021}, the gravitational-wave events GW170817 \cite{abbott_2017,abbott_2019}, GW190814 \cite{abbott_2020}, GW200105, and GW200115 \cite{abbott_2021}, as well as precise radio pulsar mass determinations such as PSR J0348+0432 \cite{antoniadis_2013}. 
\textbf{Panel (b):} Mass as a function of central density for different values of $\lambda_{\text{BL}}$. Filled circles show the maxima on $M-\mathcal{E}_c$ curves.}
\label{fig:MR_Mrho}
\end{figure}

The mass-radius (MR) relation for the given EoS is determined by numerically solving the TOV Eqs.~\eqref{tov1:eps} and \eqref{tov2:eps}) across a range of central densities, resulting in a sequence of mass-radius relationships. Fig.~\ref{fig:MR} shows the MR profiles of anisotropic stars obtained with the SLy EOS, While Fig.~\ref{fig:MR_Mrho} illustrates the variation of the stellar mass with the central energy density, within the TOV framework a perfect fluid configuration can change stability only at points where the equilibrium mass becomes stationary, that is, $\mathrm{d}M/\mathrm{d}\mathcal{E}_c = 0$. In the present work, stability is assessed using this turning point criterion. However, for anisotropic NS configurations, this condition does not necessarily guarantee full dynamical or radial oscillation stability. Moreover, satisfying standard energy and causality conditions alone is not sufficient to ensure stability in anisotropic systems, as additional instability channels, such as anisotropy-induced cracking, may arise even when conventional criteria are satisfied. A complete perturbative analysis of radial stability is therefore beyond the scope of the present work. The anisotropy parameter $\lambda_{\rm BL}$ strongly affects both the maximum mass and the corresponding stellar radius. Positive values of $\lambda_{\rm BL}$ increase the maximum mass and radius, while negative values reduce them. Table~\ref{Tab:TB1} summarizes the maximum stable mass and its associated radius for different $\lambda_{\rm BL}$, showing that negative anisotropy yields systematically lower mass–radius values than positive anisotropy. For the isotropic case ($\lambda_{\rm BL}=0.0$), the maximum stable mass is $2.051\,M_{\odot}$ with a corresponding radius of $9.850\,\mathrm{km}$.

The degree of anisotropy in NSs can be constrained using multi-messenger observations, including gravitational-wave detections, NICER measurements, and X-ray studies. The most massive and fastest rotating neutron star in the Milky Way disk, $\mathrm{PSR\ J0952\!-\!0607}$, has a measured mass of $M = 2.35 \pm 0.17\,M_\odot$ \cite{Romani_2022}, providing a stringent benchmark on the allowed anisotropy. We compare the computed mass–radius sequences with observational constraints from NICER observations of PSR~J0030+0451 \cite{riley_2019nicer,miller_2019psr} and PSR~J0740+6620 \cite{miller_2021}, gravitational-wave events GW170817 \cite{abbott_2017,abbott_2019}, GW190814 \cite{abbott_2020}, GW200105, and GW200115 \cite{abbott_2021}, as well as precise radio pulsar mass measurements such as PSR~J0348+0432 \cite{antoniadis_2013}. 

Our results are broadly consistent with these data for an appropriate range of anisotropies. In particular, GW190814, which raised the possibility that its secondary is either the lightest black hole or the most massive neutron star \cite{abbott_2020}, provides a key test of the model's high-mass regime. Consistency with the spectral EOS and universal relations inferred from GW170817 (90\% credible interval) suggests $-2 \lesssim \lambda_{\rm BL} \lesssim +2$. Moreover, the NICER measurement of PSR~J0740+6620 supports values from isotropy up to $\lambda_{\rm BL} \approx +2$, indicating that moderate positive anisotropy yields masses most compatible with current observations.

\subsection{Moment of inertia}\label{sec:MOI}
\begin{figure}[htbp]
    \centering
    % First figure
    \begin{subfigure}[b]{0.45\textwidth}
        \centering
        \includegraphics[width=\textwidth]{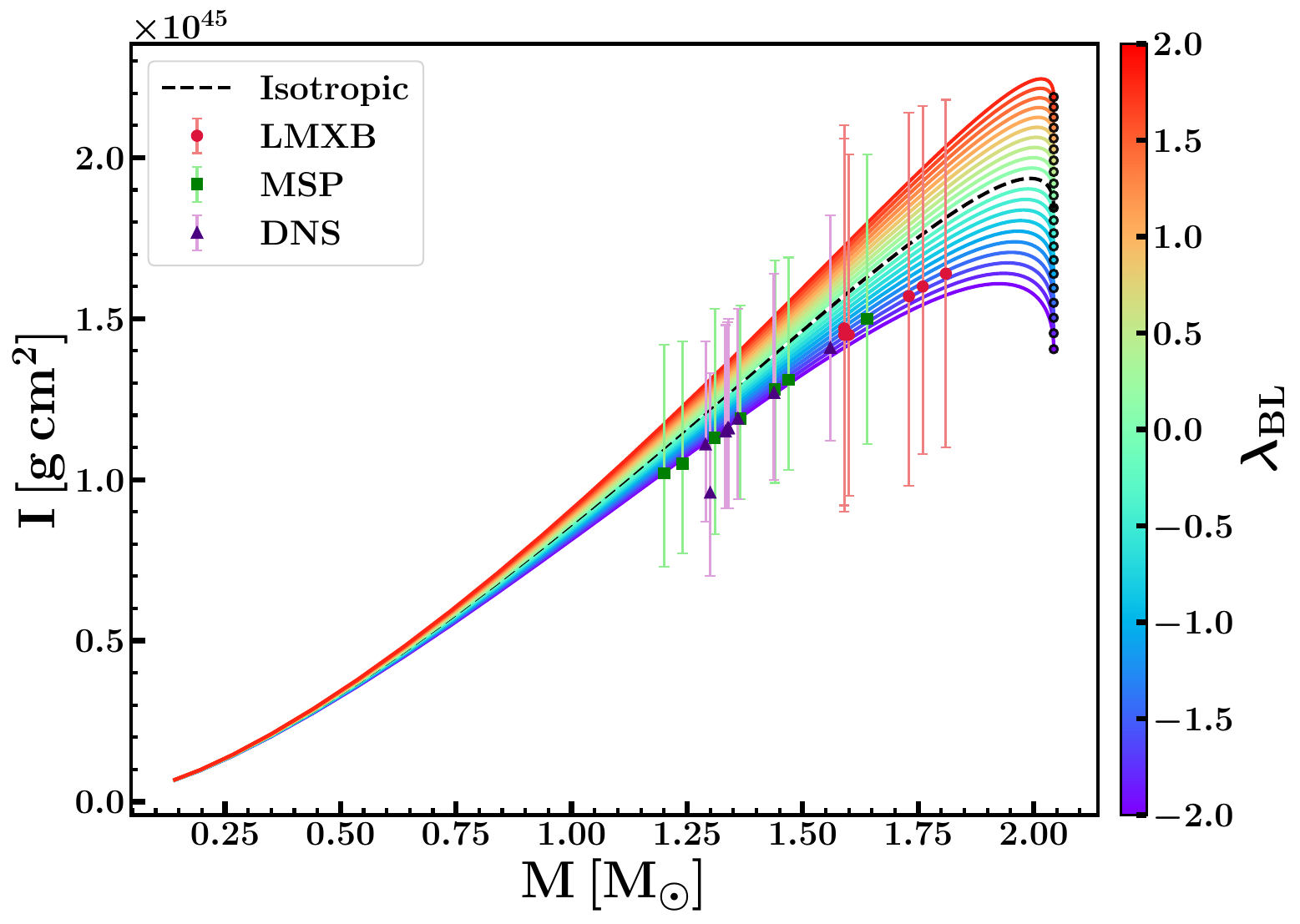}
        \caption{moment of inertia vs. mass}
        \label{fig:MI_M}
    \end{subfigure}
    \hspace{0.01\textwidth} % small gap
    % Second figure
    \begin{subfigure}[b]{0.45\textwidth}
        \centering
        \includegraphics[width=\textwidth]{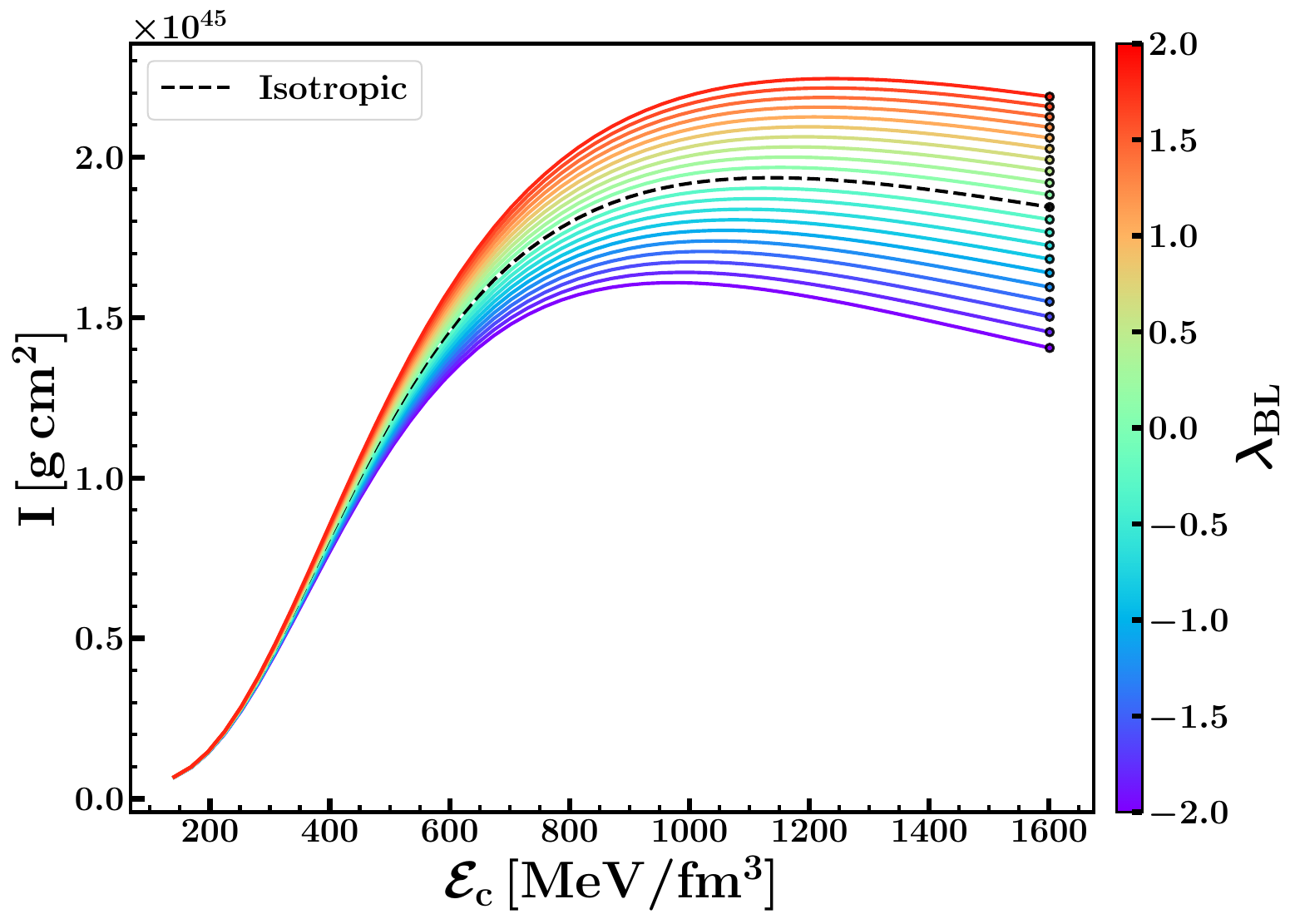}
        \caption{moment of inertia vs. central density}
        \label{fig:MI_rho}
    \end{subfigure}
    \caption{
\textbf{Panel (a):} Moment of inertia of  NSs as a function of mass for different values of the anisotropy parameter $\lambda_{\text{BL}}$. For each curve, the mass is fixed to the corresponding configuration of the isotropic case ($\lambda_{\text{BL}}=0$) while varying the parameter $\lambda_{\text{BL}}$. The error bars represent observational constraints from pulsar analyses of double NS systems (DNS), low-mass X-ray binaries (LMXB), and millisecond pulsars (MSP), as reported in Ref.~\cite{kumar_2019}. Filled circles denote the maximum moment of inertia corresponding to the maximum stable mass for the isotropic case ($\lambda_{\text{BL}}=0$) under different values of $\lambda_{\text{BL}}$.
\textbf{Panel (b):} Moment of inertia of neutron stars as a function of central energy density for different values of $\lambda_{\text{BL}}$. Here, the central density is fixed to the corresponding isotropic configuration ($\lambda_{\text{BL}}=0$) while varying $\lambda_{\text{BL}}$.
}

 \label{fig:MI_M_MI_rho}
\end{figure}

In this section, we examine the MOI of NSs described by the SLy EoS for different values of $\lambda_{\rm BL}$. The anisotropic MOI relation given in Eq.~\eqref{eq:MOI} is employed to quantify the impact of anisotropy on the rotational properties of the star. Fig.~\ref{fig:MI_M_MI_rho} presents the resulting MOI for anisotropic configurations. In Fig.~\ref{fig:MI_M}, the stellar mass is fixed to the value obtained in the isotropic case ($\lambda_{\rm BL}=0$), and the corresponding MOI is computed for different values of $\lambda_{\rm BL}$. Fi.~\ref{fig:MI_rho} shows the variation of the MOI with central energy density for several $\lambda_{\rm BL}$ values. For reference, in the isotropic case the maximum stable mass is $M_{\max}=2.051,M_\odot$, corresponding to a central energy density of approximately $1611.97 \;\mathrm{MeV \;fm^{-3}}$.  Positive values of $\lambda_{\rm BL}$ lead to larger stellar masses and higher MOI, whereas negative values produce the opposite trend. The influence of anisotropy on the MOI is more pronounced in high-mass NSs than in low-mass ones.

As reported in Ref.~\cite{kumar_2019}, the error bars denote MOI constraints for low-mass X-ray binaries (LMXBs), millisecond pulsars (MSPs), and double neutron stars (DNSs), inferred from GW170817 using universal relations. This study considers a larger sample of NSs than Refs.~\cite{landry_2020,lim_2019,jiang_2020}, though with substantially larger uncertainties and generally lower average MOI values. Nevertheless, the $I$–$M$ relations obtained in the present work remain consistent with these observational constraints for all curves corresponding to different values of $\lambda_{\rm BL}$.

\subsection{Tidal deformability}\label{sec:sub_TD}
\begin{figure}
    \centering
    \begin{subfigure}[t]{0.48\textwidth}
        \centering
        \includegraphics[width=\textwidth]{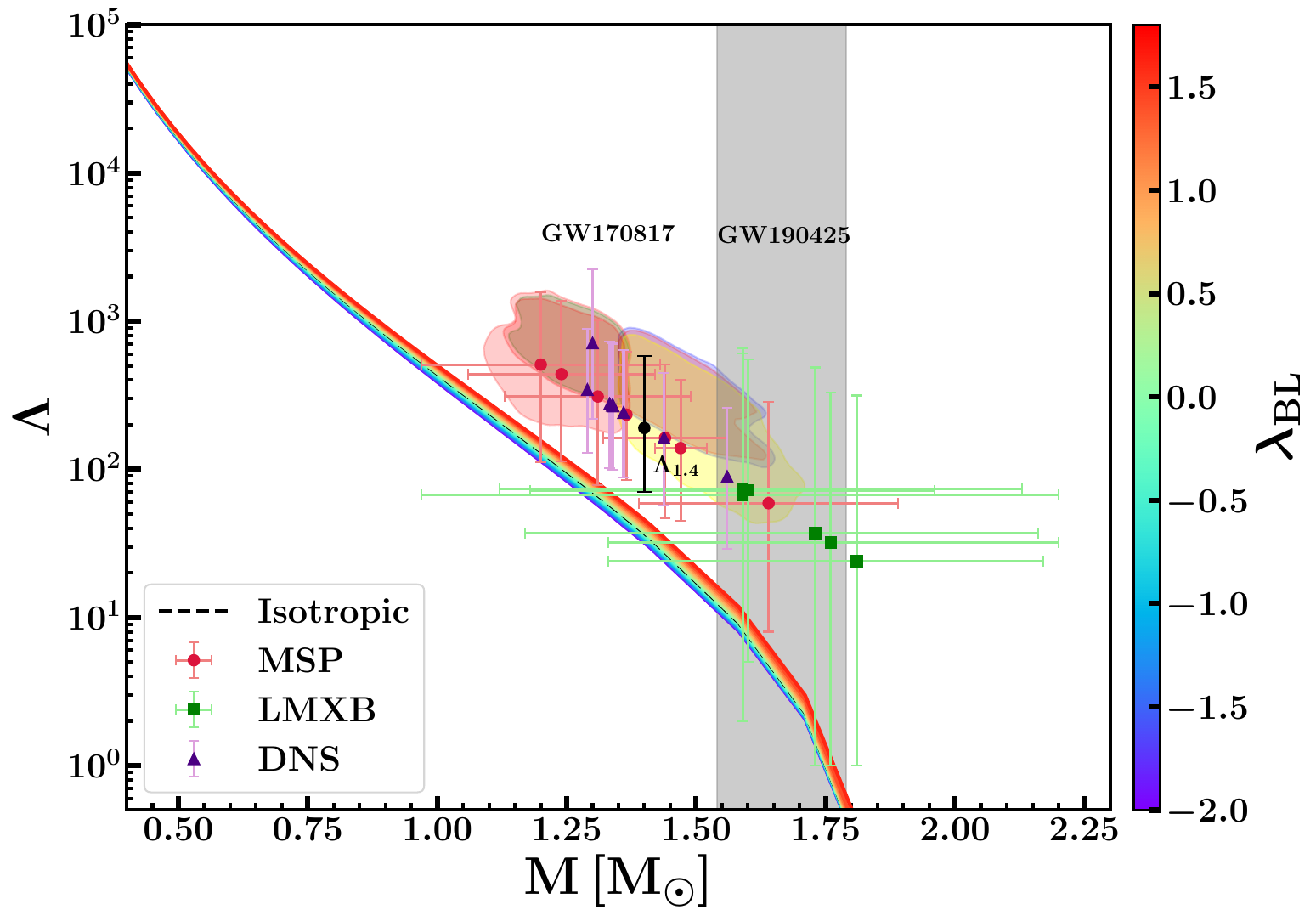}
        \caption{$\Lambda$ vs. $M$}
        \label{fig:Lambda_vs_M}
    \end{subfigure}
    \hfill
    \begin{subfigure}[t]{0.48\textwidth}
        \centering
        \includegraphics[width=\textwidth]{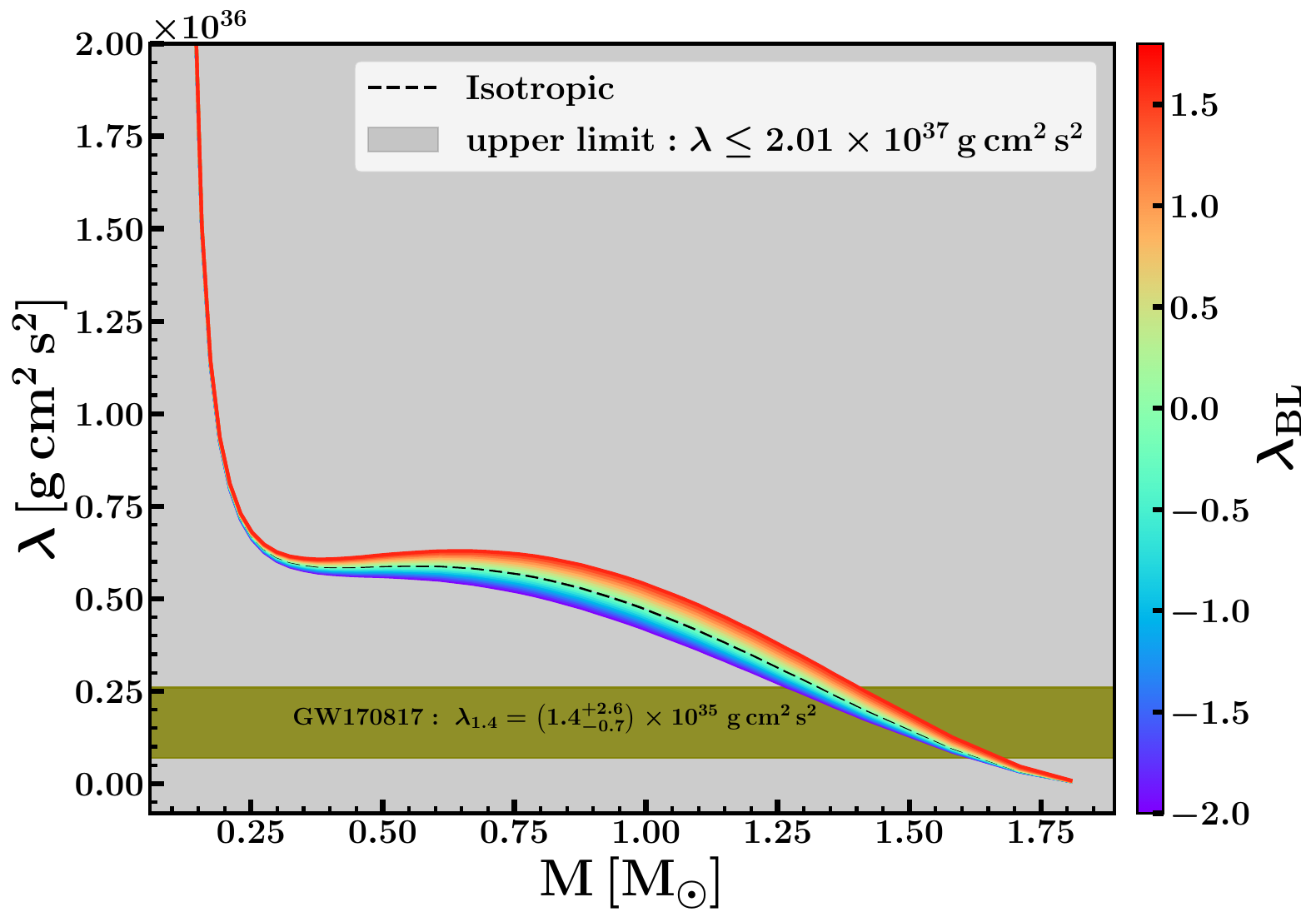}
        \caption{$\bm{\lambda}$ vs. $M$}
        \label{fig:lambda_vs_M}
    \end{subfigure}
    \vskip\baselineskip
    \begin{subfigure}[t]{0.48\textwidth}
        \centering
        \includegraphics[width=\textwidth]{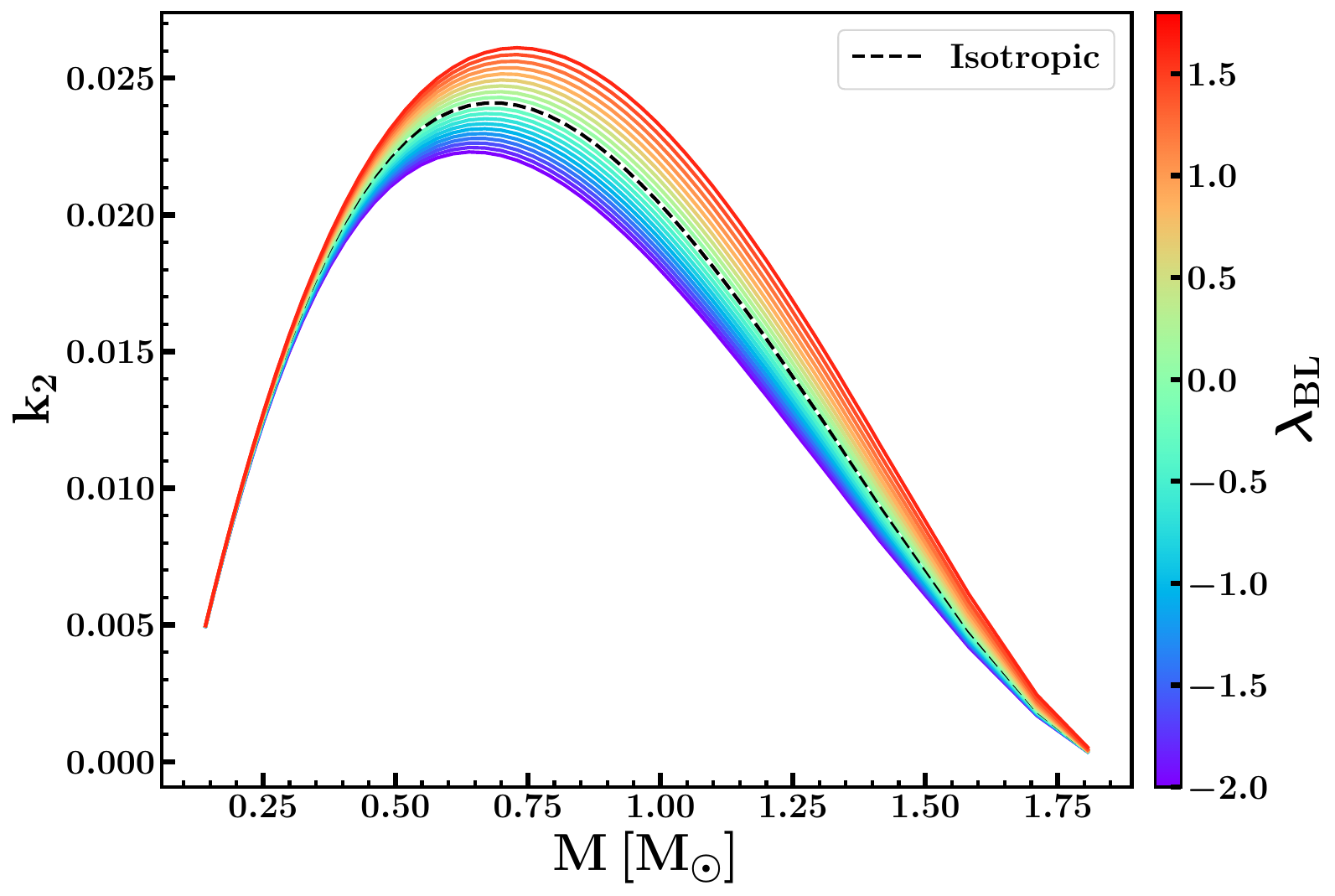}
        \caption{$k_2$ vs. $M$}
        \label{fig:k2_vs_M}
    \end{subfigure}
    \hfill
    \begin{subfigure}[t]{0.48\textwidth}
        \centering
        \includegraphics[width=\textwidth]{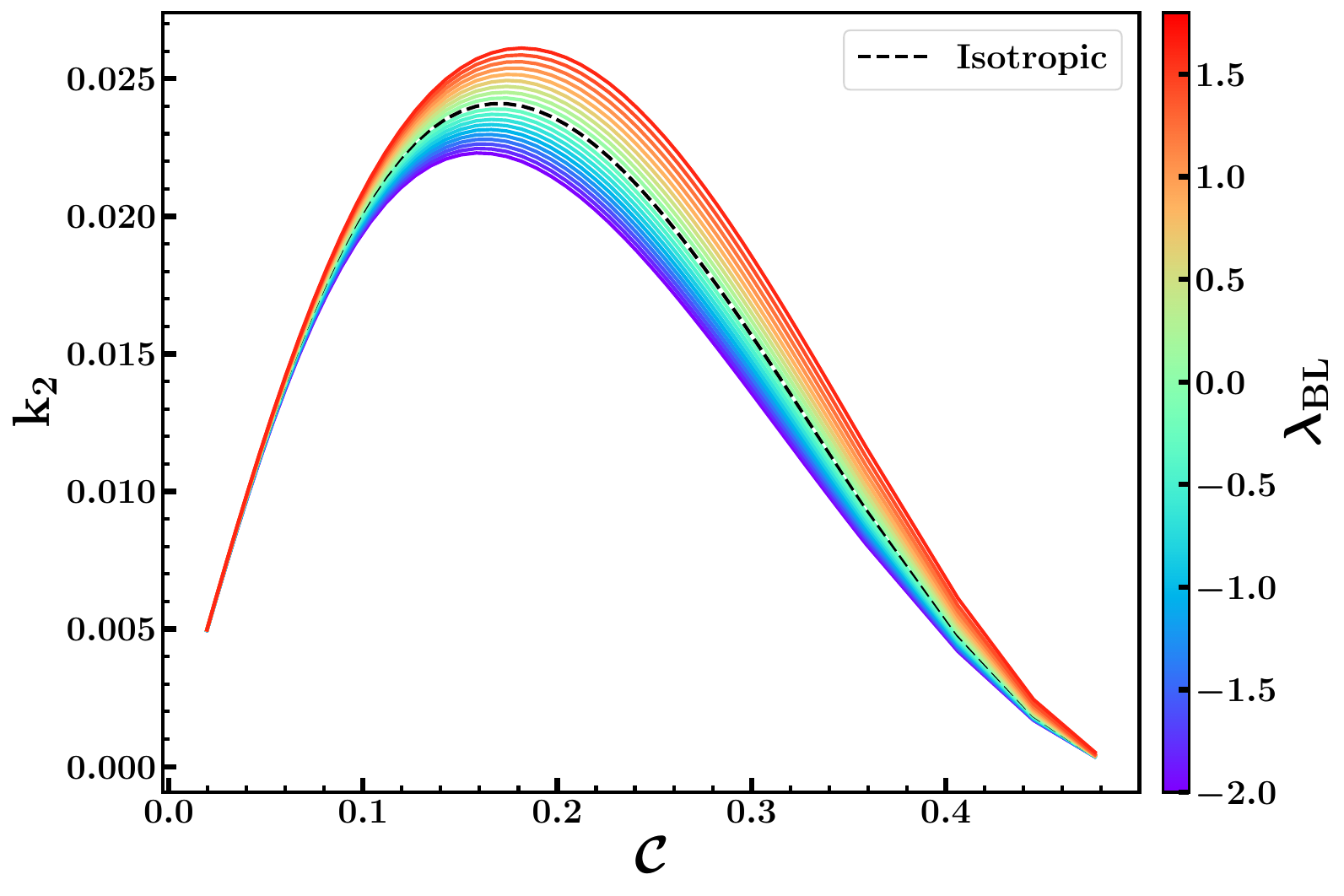}
        \caption{$k_2$ vs. $\mathcal{C}$}
        \label{fig:k2_vs_C}
    \end{subfigure}
    \caption{Tidal properties of NSs calculated for anisotropy parameters 
$\lambda_{\rm BL} \in[-2.0 ,+2.0]$ using the SLy EoS. 
\textbf{Upper panels:} (a) dimensionless tidal deformability $\Lambda$ versus mass, error bars obtained from the tidal deformability of LMXB, MSP, and DNS based on GW170817 data with universal relations are shown with observational constraints \cite{kumar_2019}. The mass range from GW190425 is shown by the gray vertical band, and for GW170817, various posteriors correlate to different methodologies and confidence levels \cite{abbott_2017,abbott_2018,abbott_2019,abbott_2020gw190425}. 
(b) tidal deformability $\bm{\lambda}$ versus mass. The gray region depicts the 90\% confidence upper limit on $\Lambda$ from X-ray and LIGO data for a binary of two nonrotating $1.4\,M_\odot$ NSs at 50 Mpc, with no tidal phase shift. The olive horizontal band represents the observational constraint from GW170817 for a $1.4\,M_\odot$ NS \cite{flanagan_2008,hinderer_2008}.
\textbf{Lower panels:} (c) dimensionless tidal Love number $k_2$ versus mass, 
(d) $k_2$ versus compactness $C$.}
\label{fig:TD_all}
\end{figure}

GWs from binary NS mergers have been observed in several events \cite{abbott_2017,abbott_2018,abbott_2019,abbott_2020}. Information about the internal structure and EoS of NSs can be extracted from the tidal deformability, encoded in the inspiral waveform. Figure~\ref{fig:TD_all} presents the numerical results for tidal properties corresponding to different values of the anisotropy parameter $\lambda_{\rm BL}$, obtained from Eq.~\eqref{Lambda}. The dependence of the dimensionless tidal deformability $\Lambda$ on stellar mass is shown in Fig.~\ref{fig:Lambda_vs_M}, while Fig.~\ref{fig:lambda_vs_M} displays the corresponding dimensionful tidal deformability $\bm{\lambda} = \tfrac{2}{3} k_2 R^5$. The variation of the dimensionless tidal Love number $k_2$ with mass and compactness $C$ is illustrated in Figs.~\ref{fig:k2_vs_M} and~\ref{fig:k2_vs_C}, respectively.

It should be noted that the tidal sequences shown in Fig.~\ref{fig:Lambda_vs_M}
terminate slightly before the maximum-mass configurations of the corresponding
 $M-\mathcal{E}_c$ relation. This occurs because the numerical integration
of the tidal perturbation equations becomes increasingly unstable close to the
turning point of the equilibrium sequence. Consequently, the tidal deformability
curves are shown only up to the last numerically stable configuration.

Within the considered anisotropy range, the overall magnitude of tidal deformability shows only modest variation with $\lambda_{\rm BL}$. Comparison with observational constraints indicates that the results are not in close agreement with GW170817 posteriors at different confidence levels and show some tension with DNS estimates. However, partial consistency is found with constraints from LMXBs and MSPs. The computed dimensional tidal deformability $\lambda$ is compatible with the limits predicted by Flanagan and Hinderer \cite{flanagan_2008,hinderer_2008}, as well as with GW170817 constraints for a $1.4\,M_\odot$ neutron star. Our anisotropic models remain consistent with these bounds for stellar masses in the range $1.27$--$1.59\,M_\odot$. The behavior of the tidal Love number indicates a systematic dependence on mass and compactness, while anisotropy significantly affects its overall magnitude.

\subsection{Surface and interior curvatures}\label{sec:sur_curv}
\begin{figure}
    \centering
    % First figure
    \begin{subfigure}[b]{0.45\textwidth}
        \centering
        \includegraphics[width=\textwidth]{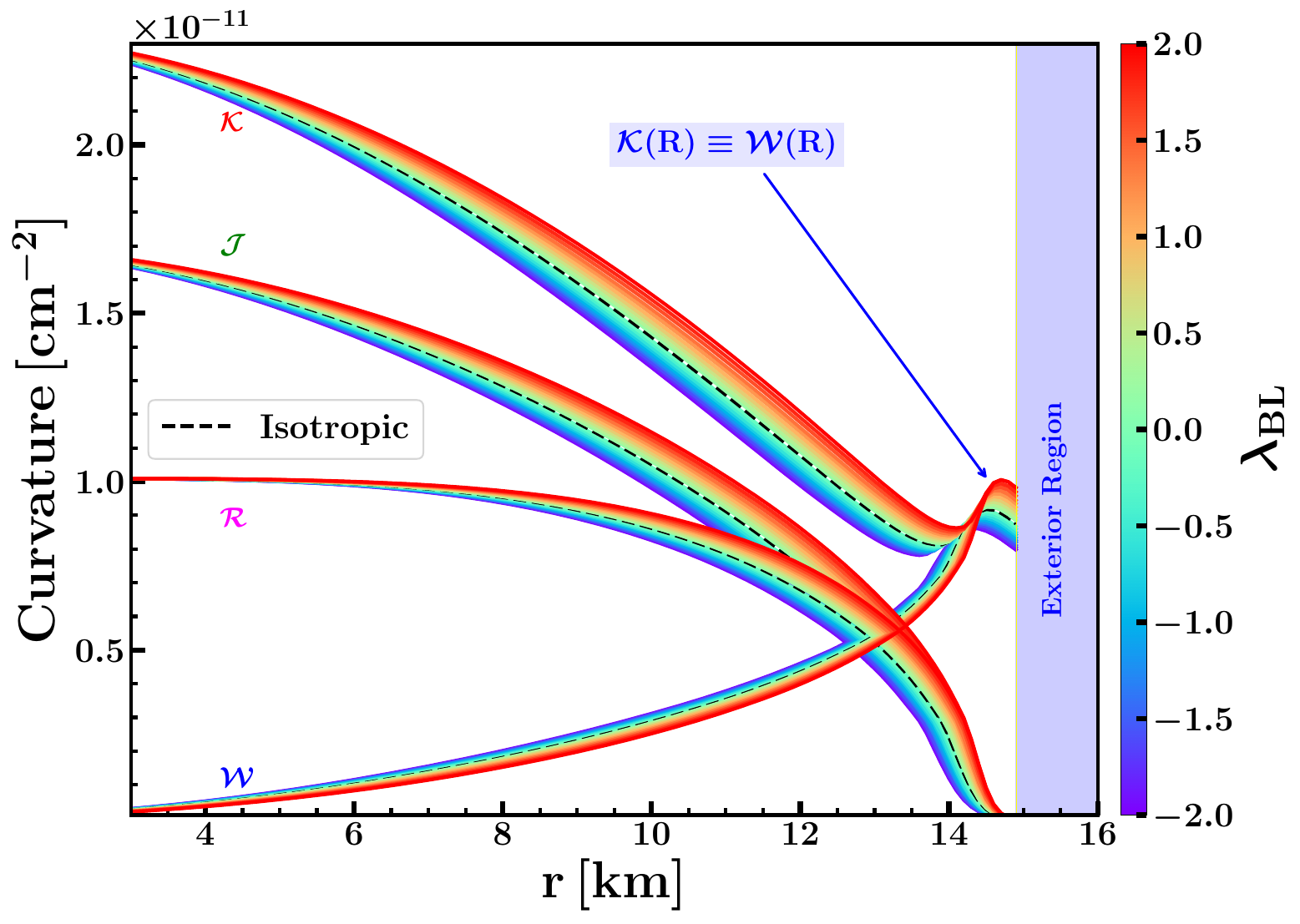}
        \caption{interior curvatures}
        \label{fig:IC}
    \end{subfigure}
    \hspace{0.01\textwidth} % small gap
    % Second figure
    \begin{subfigure}[b]{0.45\textwidth}
        \centering
        \includegraphics[width=\textwidth]{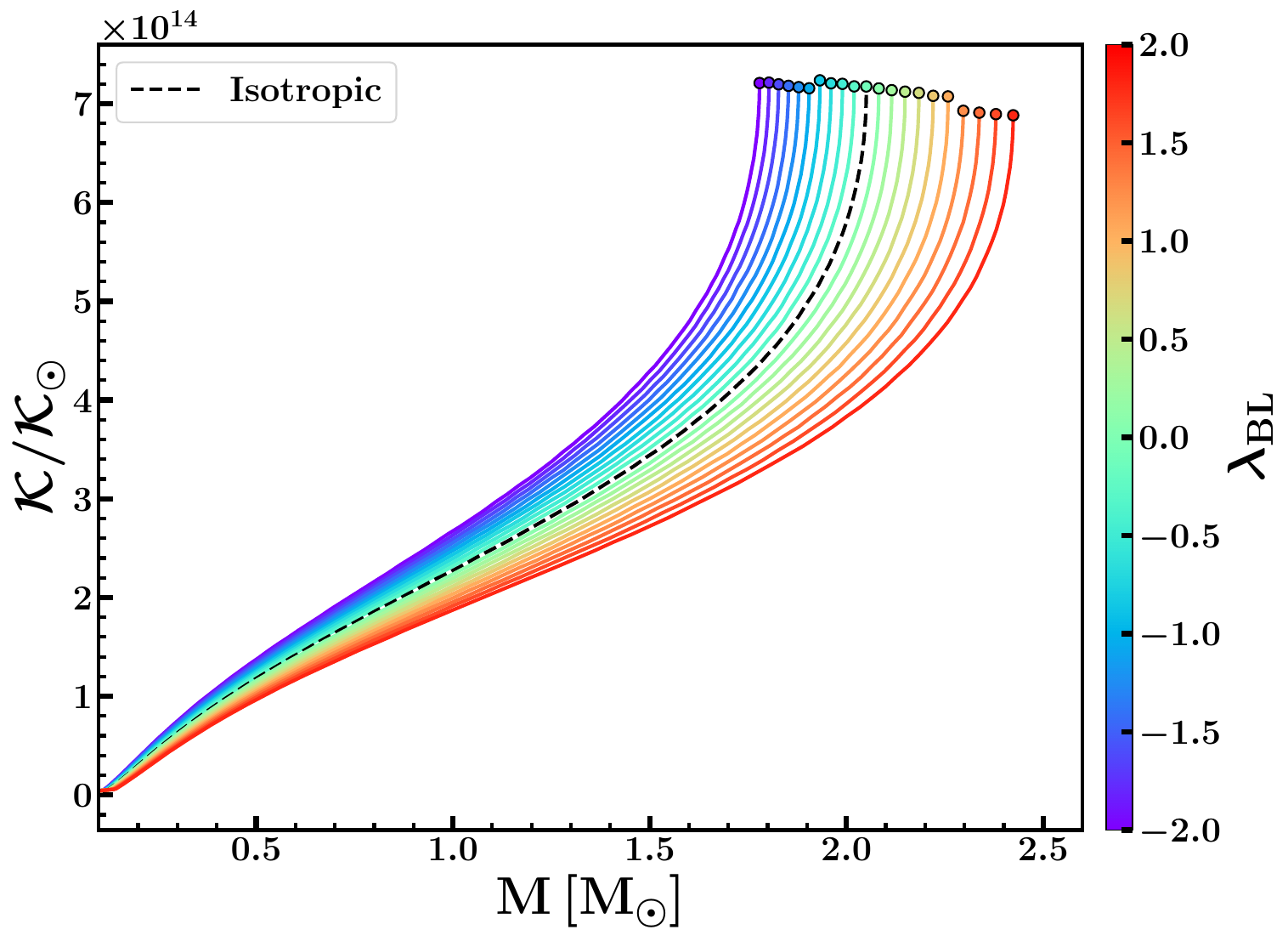}
        \caption{surface curvatures (${\cal{K}}(R)/{\cal{K}}_\odot$)}
        \label{fig:SC}
    \end{subfigure}
    \caption{\textbf{Panel (a)} Radial profiles of the $\cal{J}$, $\cal{R}$, $\cal{K}$, and $\cal{W}$ for anisotropic NSs with parameter range $-2.0 < \lambda_{\mathrm{BL}} < +2.0$ for the SLy EoS. 
    \textbf{Panel (b)} SC as a function of stellar mass, where the solar value is $\mathcal{K}_{\odot} = 4\sqrt{3}\,M_{\odot}/R_{\odot}^{3} \approx 3.06 \times 10^{-27}\,\mathrm{cm}^{-2}$. Filled circles indicate the maximum surface curvature corresponding to the $M-\mathcal{E}_c$ curves.}
    \label{fig:IC_SC}
\end{figure}

In this analysis, we evaluate the radial behavior of the curvature invariants ${\cal R}$, ${\cal J}$, ${\cal K}$, and ${\cal W}$ for NSs described by the SLy EoS, considering different values of the anisotropy parameter $\lambda_{\rm BL}$. Fig.~\ref{fig:IC} displays the corresponding curvature profiles from the stellar center to the surface. Although their magnitudes and radial trends differ across the core, crust, and exterior regions, all curvature measures attain comparable values near the crust–core boundary. A comparison across different anisotropy strengths shows that ${\cal R}$, ${\cal J}$, and ${\cal K}$ respond similarly, with their magnitudes varying systematically as $\lambda_{\rm BL}$ changes, reflecting their direct dependence on anisotropic stresses and the matter distribution inside the star.

The quantitative magnitude of the curvature invariants is expected to depend on the underlying EoS, since the EoS determines the pressure–density relation entering the TOV equations and therefore fixes the compactness and internal density profile of the star. In general, softer EoSs produce more compact configurations with higher central densities, which enhance the central curvature, whereas stiffer EoSs lead to larger radii and lower curvature values. Similar behavior has been reported in previous studies, where softer EoSs were found to generate larger surface and interior curvature due to increased compactness, although the qualitative radial behavior of the curvature profiles remains largely unchanged across realistic EoSs~\cite{eksi_2014,das_2023,das_2022,ghosh_2026}.

In contrast, ${\cal W}$ exhibits a weaker sensitivity to anisotropy, consistent with its role as a measure of the free tidal field rather than local matter concentration. At the stellar center ($r \simeq 0$), ${\cal K}$ attains the largest value, while ${\cal W}$ shows the opposite trend. Both ${\cal J}$ and ${\cal K}$ peak at the center, whereas ${\cal W}$ behaves differently. As expected, ${\cal R}$ and ${\cal J}$ vanish near the surface, while ${\cal K}$ and ${\cal W}$ increase within the crust, reaching a local maximum before approaching $4\sqrt{3}GM/R^3c^2$ at the surface.

The surface curvature (SC) increases monotonically from the stellar center to the surface, reflecting the outward strengthening of the gravitational field. Figure~\ref{fig:SC} shows the ratio ${\cal K}(R)/{\cal K}_{\odot}$ as a function of neutron star mass for different values of $\lambda_{\rm BL}$. Anisotropy has a marked effect on SC, with distinct $\lambda_{\rm BL}$ values producing different curvature levels. The compactness $\eta = 2M/R$ increases from $0.560$ to $0.678$ as $\lambda_{\rm BL}$ varies from $-2.0$ to $+2.0$, directly influencing the surface curvature: larger $\eta$ corresponds to a smaller surface ${\cal K}$ and hence a reduced SC. The corresponding numerical values are summarized in Table~\ref{Tab:TB1}.  

The curvature inside NSs is up to $10^{14}$ times stronger than that around the Sun. While ${\cal K}(r)$ dominates in the dense core, ${\cal W}(r)$ grows outward from zero as ${\cal W}(r)\propto r^{2}$ and becomes dominant near the crust. Although curvature invariants are not directly observable quantities, 
their behavior is tightly correlated with measurable stellar properties 
such as compactness, mass–radius relations, tidal deformability, and surface redshift. 
In particular, the strong curvature near the crust–core interface coincides with regions where the EoS is comparatively 
well constrained by nuclear experiments. 
Hence, deviations from GR would most likely manifest indirectly through crust-related observables such as glitches, 
cooling behavior, or surface emission properties. These results show that anisotropy affects both the stellar structure and spacetime curvature, underscoring the importance of testing general relativity in such extreme strong-field environments \cite{eksi_2014,psaltis_2008}.

\subsection{Inhomogeneity}\label{sec:inho}
We find that spacetime curvature varies sensitively with both anisotropy and matter distribution across the neutron star interior. While the equation of state, often formulated in flat spacetime, is commonly used to describe neutron star structure, our results show that curvature provides essential additional information in the strong-field regime. In particular, the coupled interplay between curvature and anisotropy is crucial for a realistic description of the stellar interior.

The Weyl curvature represents the free tidal component of the gravitational field and therefore carries information about how matter is spatially distributed inside the star. In spherical symmetry, a perfectly homogeneous matter distribution corresponds to vanishing Weyl curvature, whereas density gradients or anisotropic stresses generate nonzero Weyl curvature. This makes the Weyl scalar a natural indicator of internal inhomogeneity, complementing the information from the equation of state. Since anisotropic pressure modifies the local force balance and redistributes matter inside the star, it is expected to be directly linked with the development of curvature-driven inhomogeneity.

To quantify this coupling, we analyze the tidal contribution encoded in ${\cal W}$, which is sensitive to inhomogeneities in the energy–momentum distribution. We introduce an inhomogeneity parameter $\zeta$, defined as the ratio of the Weyl curvature to the anisotropy factor $\sigma$,
\begin{equation}
\zeta(r) := \frac{{\cal W}(r)}{\sigma(r)} ,
\end{equation}
where $\sigma(r)$ denotes the difference between tangential and radial pressures. The quantity $\zeta$ can be interpreted as a local response measure that quantifies how efficiently anisotropic stresses generate tidal curvature. In other words, it represents the amount of Weyl curvature produced per unit anisotropy and therefore provides a diagnostic of the coupling between anisotropic stresses and gravitational inhomogeneity.

A subtle but important point concerns the isotropic limit. When the pressure becomes isotropic, $\sigma \to 0$, the ratio defining $\zeta$ becomes formally indeterminate. This does not indicate any physical inconsistency; rather, it reflects that $\zeta$ is meaningful only in the presence of anisotropic stresses. In isotropic configurations, inhomogeneity is governed solely by density gradients, and the Weyl curvature itself must be used directly to characterize the internal structure. For this reason, $\zeta$ should be regarded as an anisotropy-sensitive diagnostic rather than a universal invariant valid in all cases.
 A nonzero value of $\zeta$ indicates the presence of curvature-driven inhomogeneity associated with anisotropic stresses.

Special care is required near the stellar center ($r \lesssim 1\,\mathrm{cm}$), where both ${\cal W}(r)$ and $\sigma(r)$ approach zero, leading to potential numerical cancellation errors. To address this, we employ series expansions of ${\cal W}(r)$ and $\sigma(r)$ about the origin, as detailed in Appendix~\ref{APD:W_sigma}, ensuring a stable and physically consistent evaluation of $\zeta(r)$ throughout the star.

\begin{figure}
    \centering
    % First figure
    \begin{subfigure}[b]{0.45\textwidth}
        \centering
        \includegraphics[width=\textwidth]{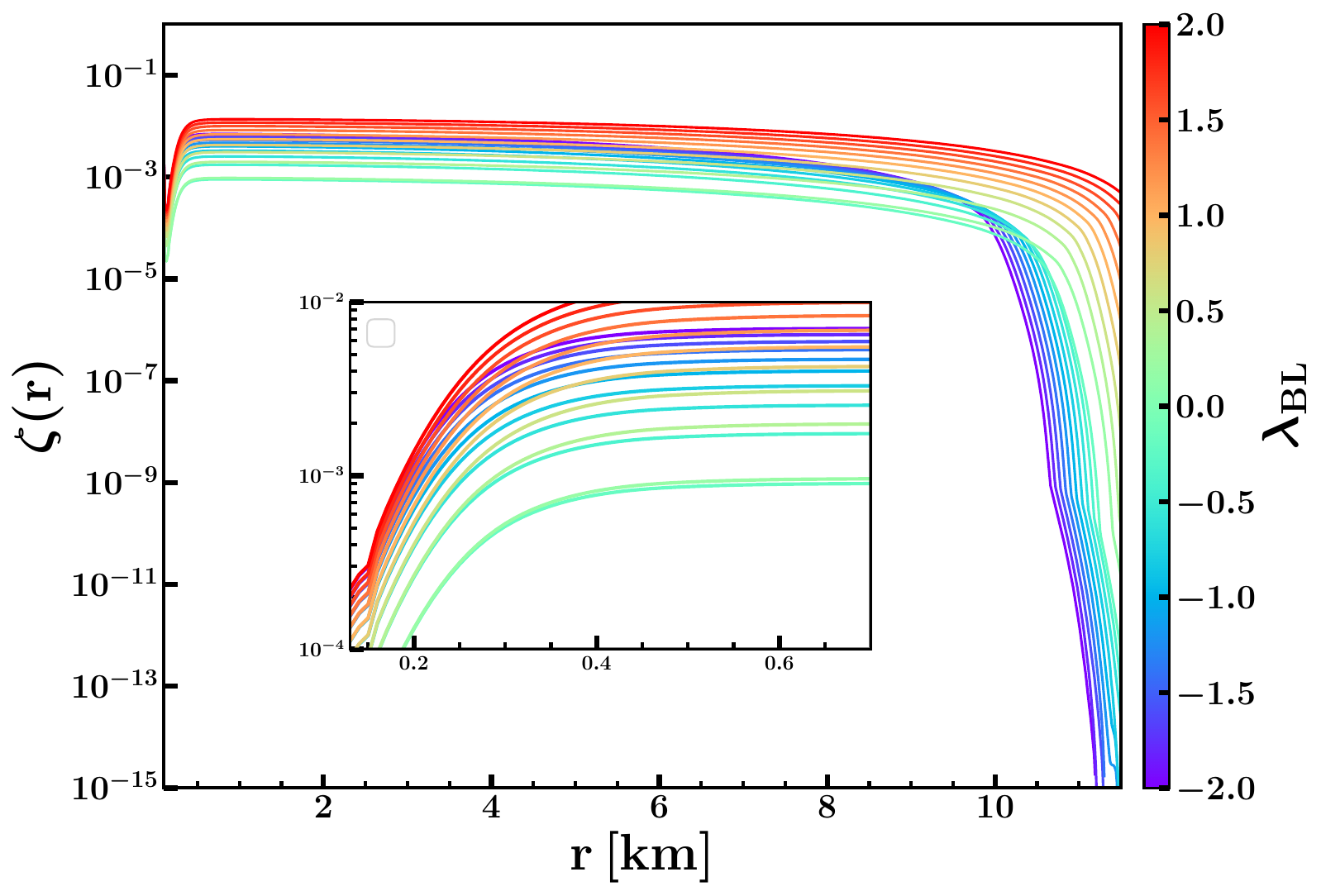}
        \caption{$\zeta(r)$ for varying $\lambda_{\mathrm{BL}}$}
        \label{fig:zeta_a}
         \end{subfigure}
    \hspace{0.01\textwidth} % small gap
    % Second figure
    \begin{subfigure}[b]{0.45\textwidth}
        \centering
        \includegraphics[width=\textwidth]{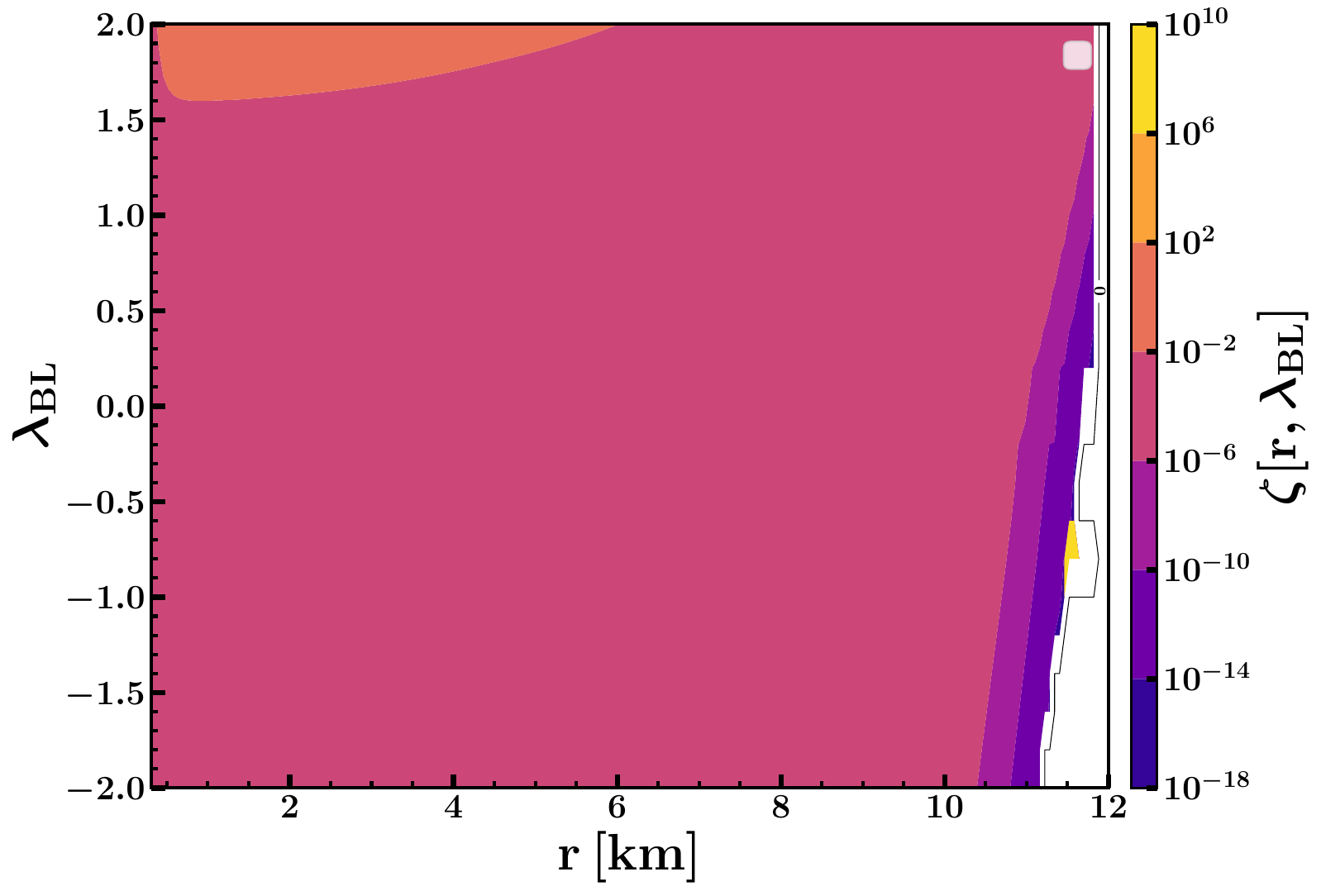}
        \caption{Contour of $\zeta$ vs. $r$ and $\lambda_{\mathrm{BL}}$}
        \label{fig:zeta_cont}
    \end{subfigure}
    \caption{\textbf{(a)} Inhomogeneity parameter $\zeta(r)$ for different values of the anisotropy parameter $\lambda_{\mathrm{BL}}$. \textbf{ (b)} Variation of $\zeta$ with $r$ and $\lambda_{\mathrm{BL}}$ shown as a contour plot.}
    \label{fig:Zeta}
\end{figure}

The behavior of $\zeta(r)$ is shown in Fig.~\ref{fig:Zeta}, including its radial profile and a contour representation in the $(r,\lambda_{\rm BL},\zeta)$ space. The inhomogeneity parameter remains very small near the center, increases gradually with radius, and approaches a nearly stable trend toward the surface. This qualitative behavior persists for all explored values of $\lambda_{\rm BL}$, while the contour plot clearly illustrates how anisotropy progressively enhances curvature-driven inhomogeneity toward the outer regions of the star.

\sisetup{
    scientific-notation = true,
    round-mode          = figures,
    round-precision     = 4,
    table-number-alignment = center,
    exponent-product    = \times,
    separate-uncertainty = true
}
% for row coloring
\sisetup{
    round-mode          = places,
    round-precision     = 3,
    table-number-alignment = center
}
\begin{table}[ht]
\centering
\caption{Numerical values of the computed stellar properties of anisotropic NSs for different values of the  ($\lambda_{\mathrm{BL}}$). 
The listed quantities include the NS mass  ($M_{\mathrm{NS}}/M_{\odot}$), radius ($R$ in km), compactness ($\eta$), normalized compactness ($\eta/\eta_{\odot} \times 10^{5}$), surface curvature ($\mathcal{K} \times 10^{-12}\,\mathrm{cm}^{-2}$), and normalized surface curvature ($\mathcal{K}/\mathcal{K}_{\odot} \times 10^{14}$).}

\label{tab:compact_objects}
\begin{tabular}{ccccccc}
\toprule
$\lambda_{\rm BL}$ 
& $M/M_\odot$ 
& $R \;(\rm km)$ 
& $\eta$ 
& \makecell{$\eta / \eta_\odot$ \\($10^{5}$)}
& \makecell{$\mathcal{K}$ \\ ($10^{-12}\; \rm cm^{-2}$)} 
& \makecell{$\mathcal{K} / \mathcal{K}_\odot$ \\ ($10^{14}$)} \\
\midrule
-2.0 & 1.779 & 9.380 & 0.560 & 1.321 & 2.205 & 7.278 \\
-1.8 & 1.803 & 9.420 & 0.565 & 1.333 & 2.207 & 7.283 \\
-1.6 & 1.828 & 9.470 & 0.570 & 1.344 & 2.202 & 7.268 \\
-1.4 & 1.853 & 9.520 & 0.575 & 1.355 & 2.197 & 7.252 \\
-1.2 & 1.879 & 9.570 & 0.580 & 1.367 & 2.193 & 7.239 \\
-1.0 & 1.905 & 9.620 & 0.585 & 1.379 & 2.189 & 7.225 \\
-0.8 & 1.933 & 9.630 & 0.593 & 1.398 & 2.214 & 7.308 \\
-0.6 & 1.961 & 9.690 & 0.598 & 1.409 & 2.205 & 7.277 \\
-0.4 & 1.990 & 9.740 & 0.603 & 1.423 & 2.203 & 7.272 \\
-0.2 & 2.020 & 9.800 & 0.609 & 1.435 & 2.196 & 7.247 \\
\rowcolor{blue!20} % highlight row
0.0  & 2.051 & 9.850 & 0.615 & 1.450 & 2.196 & 7.246 \\
0.2  & 2.082 & 9.910 & 0.621 & 1.463 & 2.189 & 7.223 \\
0.4  & 2.115 & 9.970 & 0.627 & 1.477 & 2.183 & 7.206 \\
0.6  & 2.149 & 10.03 & 0.633 & 1.492 & 2.179 & 7.191 \\
0.8  & 2.184 & 10.09 & 0.639 & 1.507 & 2.175 & 7.179 \\
1.0  & 2.221 & 10.16 & 0.646 & 1.522 & 2.167 & 7.150 \\
1.2  & 2.258 & 10.22 & 0.653 & 1.538 & 2.164 & 7.142 \\
1.4  & 2.297 & 10.35 & 0.655 & 1.545 & 2.120 & 6.995 \\
1.6  & 2.338 & 10.42 & 0.663 & 1.562 & 2.114 & 6.978 \\
1.8  & 2.380 & 10.49 & 0.670 & 1.580 & 2.109 & 6.962 \\
2.0  & 2.424 & 10.56 & 0.678 & 1.598 & 2.106 & 6.950 \\
\bottomrule
\end{tabular}
\label{Tab:TB1}
\end{table}

\subsection{Upper compactness in anisotropic neutron stars}

Rezzolla et al.~\cite{rezzolla_2025} investigated the maximum compactness achievable by neutron stars by constructing a large ensemble of equations of state consistent with nuclear theory, perturbative QCD, and current observations. They showed that, within isotropic TOV models satisfying nuclear-theory and perturbative-QCD constraints, the configuration attaining the maximum TOV mass is also the most compact, leading to an EoS-insensitive upper bound $\mathcal{C}_{\max} \lesssim 1/3$. Motivated by this result, and treating the isotropic bound as a comparative reference rather than a strict constraint, we revisit the problem within the framework of anisotropic stellar models. Using the BL prescription, we vary the anisotropy parameter $\lambda_{\mathrm{BL}}$ within the range $[-2,+2]$, noting that for $\lambda_{\mathrm{BL}} \lesssim -1.8$ the tangential pressure begins to exhibit unphysical behavior near the stellar surface; the chosen interval therefore ensures regular and physically acceptable configurations throughout the star.

\begin{figure}[H]
    \centering
    % First figure
    \begin{subfigure}[b]{0.46\textwidth}
        \centering
        \includegraphics[width=\textwidth]{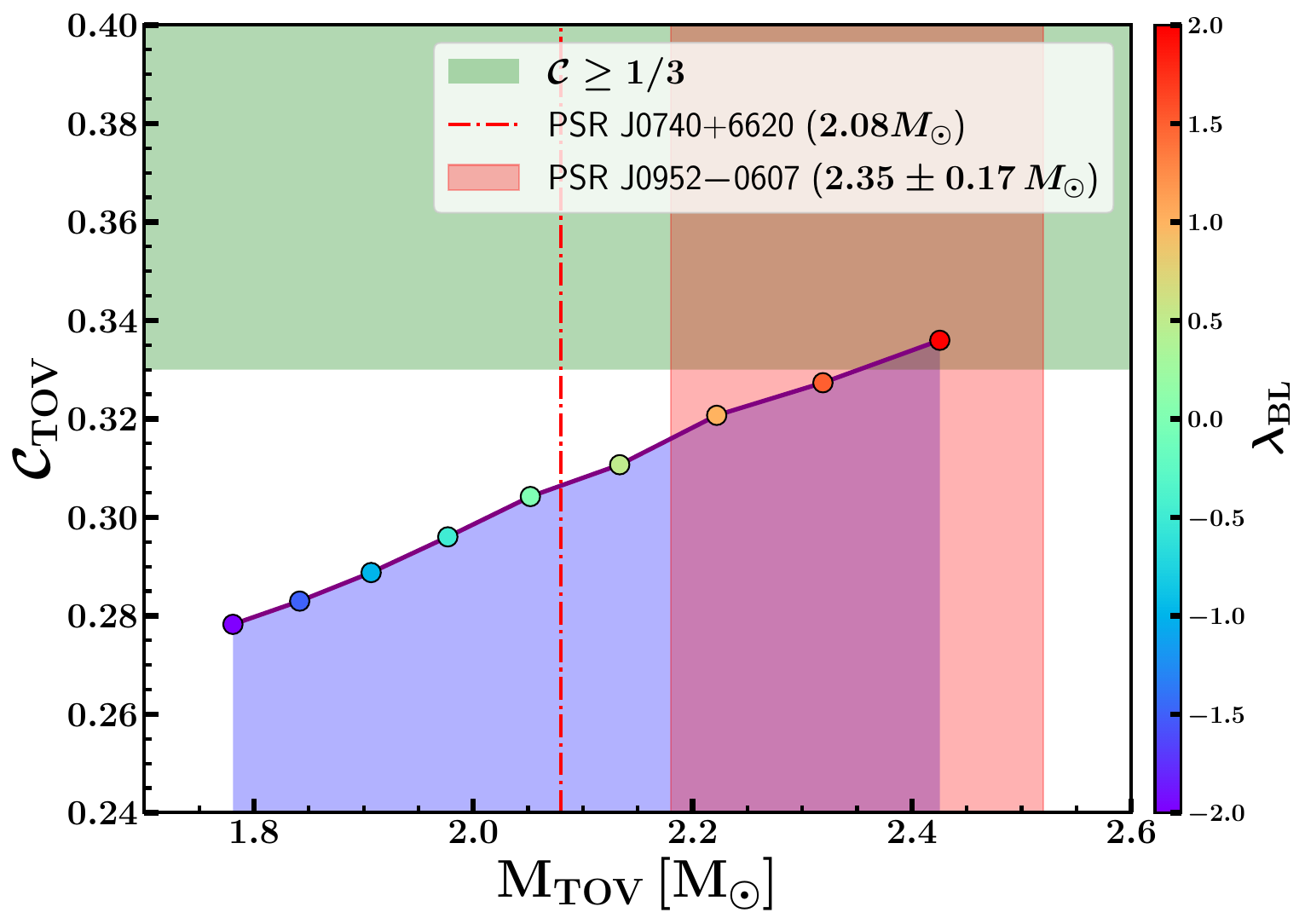}
        \caption{$\mathcal{C_{\mathrm{TOV}}} \; \rm vs\; \mathrm{M_{\mathrm{TOV}}}$}
        \label{fig:Cmax_a}
         \end{subfigure}
    \hspace{0.01\textwidth} % small gap
    % Second figure
    \begin{subfigure}[b]{0.48\textwidth}
        \centering
        \includegraphics[width=\textwidth]{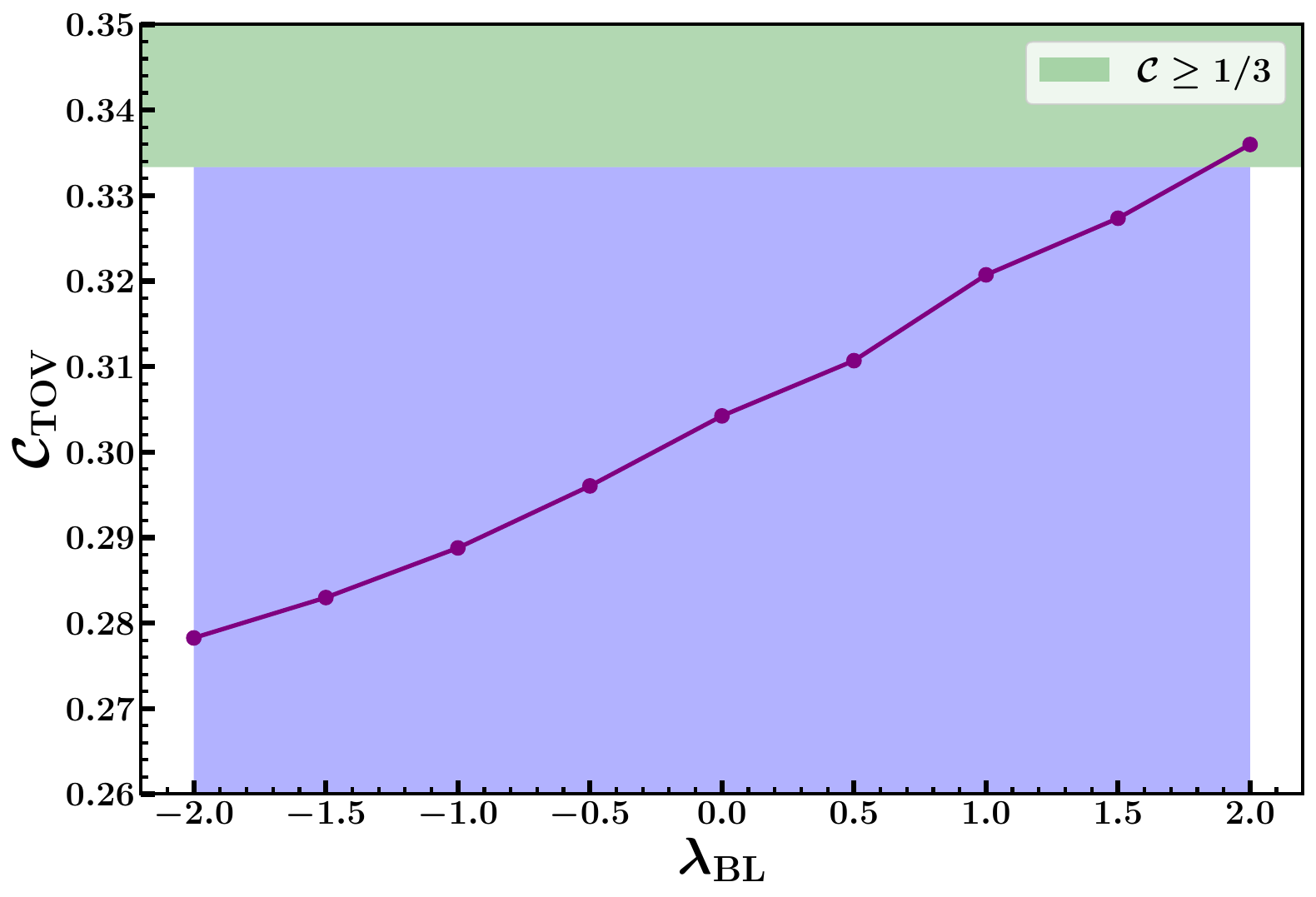}
        \caption{$\mathcal{C_{\rm TOV}}\; \rm vs \; \lambda_{\rm BL}$}
        \label{fig:Cmax_b}
    \end{subfigure}
    \caption{\textbf{(a)} $\mathcal{C}_{\mathrm{TOV}}$ versus the TOV maximum mass $\rm M_{\mathrm{TOV}}$ for $\lambda_{\mathrm{BL}} \in [-2,+2]$, with the light-red band indicating the mass of PSR~J0952$-$0607 ($2.35 \pm 0.17\,\rm M_{\odot}$)~\cite{Romani_2022} and the vertical red dotted line marking PSR~J0740$+$6620 ($2.08\,\rm M_{\odot}$)~\cite{miller_2021}. \textbf{(b)} $\mathcal{C}_{\mathrm{TOV}}$ as a function of the anisotropy parameter $\lambda_{\mathrm{BL}}$, where the horizontal axis directly spans the considered values of $\lambda_{\mathrm{BL}}$.}
    \label{fig:Cmax}
\end{figure}

Fig.~\ref{fig:Cmax} illustrates the dependence of the maximum compactness $\mathcal{C}_{\max}$ on the corresponding TOV mass $\mathrm{M}_{\mathrm{TOV}}$ for different $\lambda_{\mathrm{BL}}$. Within the physically admissible range $\lambda_{\mathrm{BL}} \in [-2.0,2.0]$, the allowed masses span $\sim 1.78$--$2.42\,M_{\odot}$, with compactness varying from $\mathcal{C}_{\max} \simeq 0.28$ to $\simeq 0.37$, remaining consistent with theoretical expectations and observed massive neutron stars.
Fig.~\ref{fig:Cmax_a} also includes observational constraints from PSR~J0952$-$0607 ($2.35 \pm 0.17\,M_{\odot}$)~\cite{Romani_2022} and PSR~J0740$+$6620 ($2.08\,M_{\odot}$)~\cite{miller_2021}. This indicates that moderate anisotropy leads to NS configurations compatible with current mass measurements, while larger magnitudes of $\lambda_{\mathrm{BL}}$ either violate physical acceptability conditions or move the maximum mass outside the observed range.

We emphasize that the compactness bound $\mathcal{C}_{\max} \lesssim 1/3$ is derived under the assumption of isotropic pressure and does not necessarily remain valid in the presence of anisotropy; in the present work, it is used solely as a benchmark for comparison.

\section{Discussion and conclusions}\label{sec:Disc}

Neutron star interiors are challenging to model due to their extreme densities and strong gravity. The microphysics of dense matter, natural phase transitions, superfluidity, and strong magnetic fields can all cause pressure anisotropy, which is often overlooked in conventional isotropic models. Incorporating anisotropy allows models to capture a broader range of structural and rotational properties, to correlate theoretical predictions with multi-messenger observations, and to better represent the correlation between matter distribution and spacetime curvature in the strong-field regime.

We emphasize that the use of the SLy EoS in this work serves mainly as a representative choice; based on previous curvature analyses with different realistic EoSs, we expect that adopting softer or stiffer models would primarily rescale the magnitude of the curvature invariants without altering the qualitative conclusions regarding anisotropy effects.

In this study, we employed the BL model to investigate the properties of anisotropic NSs. The results obtained suggest that the degree of anisotropy influences macroscopic quantities, since the hydrostatic balance is augmented by an additional term arising from the pressure difference between $P_\perp$ and $P_r$. This impact, which depends on the model, changes the star's equilibrium configuration and affects its overall structure.

In Sec.~\ref{sec:sub_mr}, we provided our findings about the $M$-$R$ relationship of NSs with varied degrees of anisotropy. In the isotropic $\lambda_{\rm BL}=0$ setting, without anisotropy, the highest mass is around $2\; \Ms$ with a radius of approximately $9.85~\mathrm{km}$, which is in agreement with observations of PSR J0348+0432. When the $\lambda_{\rm BL}$ varies from $-2.0$ to $+2.0$, $M$ and $R$ increase.  Positive anisotropy enables NSs to obtain masses higher than  $2\;\Ms$, with a maximum mass of approximately $2.4\;\Ms$ and a radius of around $10.5~\mathrm{km}$ for $\lambda_{\rm BL}=2.0$. Over this range of $\lambda_{\rm BL}$, the star's compactness changes significantly, increasing by nearly 21\% from the least to the maximum anisotropy value. We used the slow-rotation approach for anisotropic NSs, which is discussed in Sec.~\ref{sec:sub_MI}, in order to find out the moment of inertia. We used this approach to determine the MOI for the specified range of anisotropy parameters. The results we obtained show that the variation of the MOI with mass correlates well with the observational constraints from DNS, LMXB, and MSP observations across the full range of $\lambda_{\rm BL}$. In Sec.\ref{sec:sub_TD}, we present the dimensionless and dimensional tidal deformability, together with the tidal Love number $k_2$, as functions of mass and compactness. The tidal deformability $\Lambda$ changes only marginally when the $\lambda_{\rm BL}$ value changes, and it varies considerably less compared to other stellar properties. Our results are compatible with the 90\% C.L.\ from GW170817, $1.4\,M_{\odot}$ NS: $\lambda_{1.4} = \left(1.4^{+2.6}_{-0.7}\right)\times 10^{35}\ \mathrm{g\,cm^{2}\,s^{2}}$, for NS masses between $1.27$ and $1.59,M\odot$; yet they do not match with the observations of MSPs, LMXBs, DNSs, and GW170817.

In Secs.~\ref{sec:sur_curv}--\ref{sec:inho}, we expanded our investigation beyond conventional observables by examining the curvature invariants $\mathcal{R}$, $\mathcal{J}$, $\mathcal{K}$, and $\mathcal{W}$ to explore the interior gravitational field of anisotropic NSs. These geometric dimensions illustrate physical properties that are not accessible by mass, radius, or tidal deformability alone.

The curvature scalars exhibit different strengths and spatial trends throughout the stellar interior, though they converge near the crust-core boundary. The Ricci-related invariants $\mathcal{R}$, $\mathcal{J}$, and $\mathcal{K}$ respond coherently to perturbations in the anisotropy parameter $\lambda_{\rm BL}$, demonstrating their dependency on matter distribution and pressure anisotropy. The Weyl curvature $\mathcal{W}$, which characterizes the free tidal field, has a far reduced sensitivity to anisotropy, indicating its matter-independent origin.

At the stellar center, $\mathcal{K}$ has the highest value among all curvature invariants, reflecting the extreme curvature of the dense core. The Weyl scalar expands as $\mathcal{W}\propto r^{2}$, eventually dominating the curvature behavior near the crust, whereas $\mathcal{R}$ and $\mathcal{J}$ vanish at the surface. Meanwhile, $\mathcal{K}$ and $\mathcal{W}$ reach their Schwarzschild limits, signifying the transition from matter-dominated to curvature-dominated regions. Surface curvature decreases as compactness $\eta = 2M/R$ expands with $\lambda_{\rm BL}$, suggesting that anisotropy has a direct impact on the gravitational field gradient at the boundary.

Spacetime curvature in NS interiors is nearly $10^{14}$ times higher than solar values, underscoring the importance of curvature-based diagnostics for studying strong-field gravity. Yet, it remains difficult to fully separate anisotropy from dense-matter microphysics in such extreme conditions.

In addition, the bounds of NS compactness have been investigated in anisotropic configurations. Recent isotropic studies identify an approximate statistical upper limit around $\mathcal{C}_{\max} \sim 1/3$ under nuclear-theory and perturbative-QCD constraints. The present study extends this evaluation to anisotropic models described by the BL prescription. The analysis is restricted to $\lambda_{\mathrm{BL}} \in [-2,+2]$, excluding more negative values for which the tangential pressure develops unphysical behavior near the stellar surface. Pressure anisotropy significantly influences both the maximum mass $M_{\mathrm{TOV}}$ and the associated compactness $\mathcal{C}_{\max}$. The maximum masses for $\lambda_{\mathrm{BL}} \in [-2,+2]$ are in the range $1.78$--$2.42\,M_{\odot}$, with compactness $\mathcal{C}_{\max} \approx 0.28$--$0.37$. A comparison with current high-mass NS measurements shows that moderate anisotropy remains compatible with observations, restricting the viable parameter space for anisotropy in NS interiors.

To be thorough, we computed the anisotropic stellar properties using the QL anisotropy model with the same input parameters as the BL setting. The corresponding $M$-$R$ relations, $\mathcal{C}$, and curvature invariants are discussed in Appendix~\ref{APD:QL}. The QL model shows a much higher sensitivity to anisotropy: for $\lambda_{\mathrm{BL}}\in[-2,+2]$, the BL model provides maximum masses in the range $1.77$--$2.42\,M_{\odot}$, while the QL model results in a much wider spectrum of $1.12$--$4.56\,M_{\odot}$. This higher sensitivity means that, compared with the BL model, QL anisotropy notably amplifies the impact of anisotropic stresses on overall stellar properties.

\appendix

\section{Quasi-local anisotropy model}\label{APD:QL}
\begin{figure}
    \centering
    \begin{subfigure}[t]{0.48\textwidth}
        \centering
        \includegraphics[width=\textwidth]{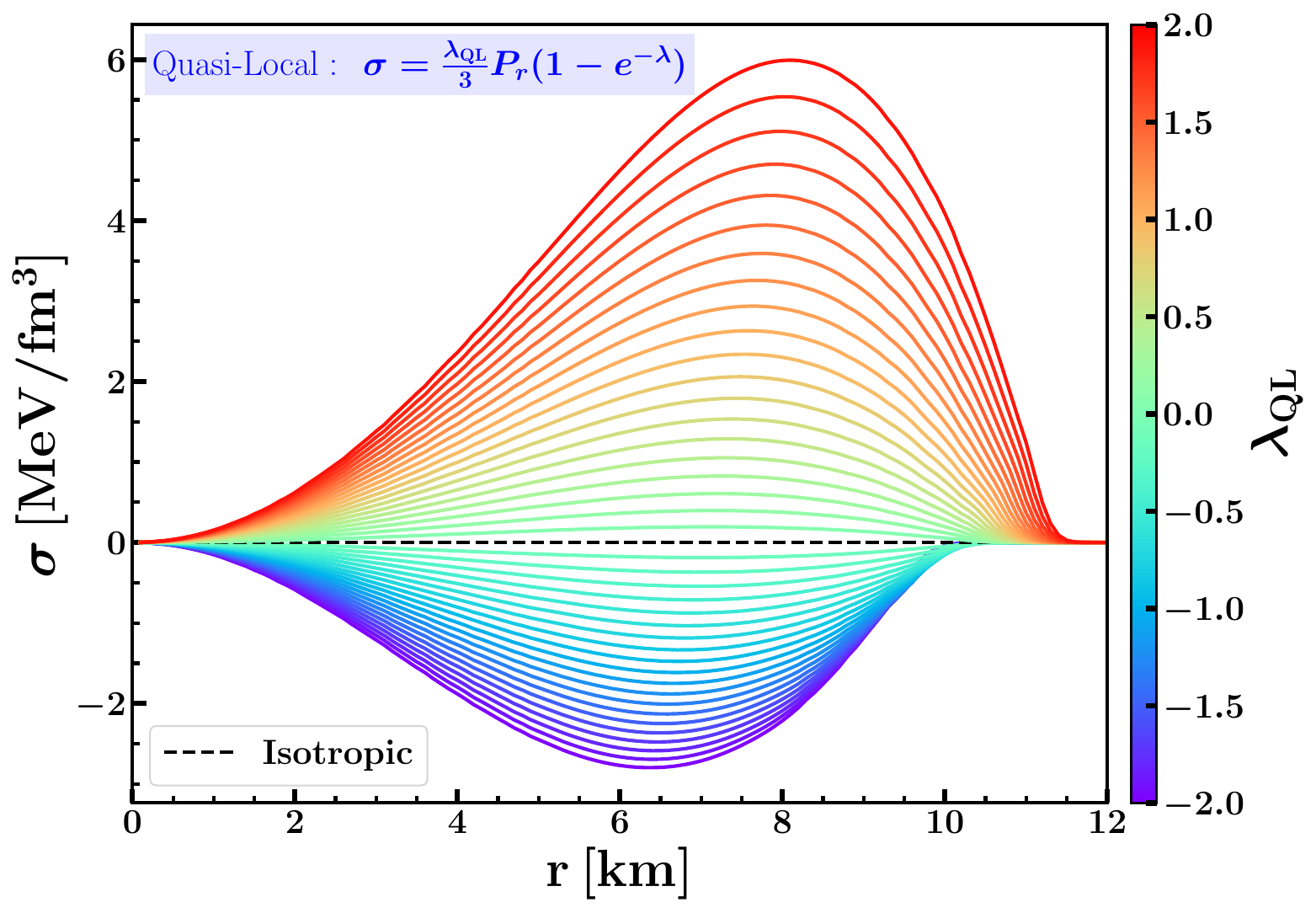}
        \caption{}
        \label{fig:QL_sigma}
    \end{subfigure}
    \hfill
    \begin{subfigure}[t]{0.48\textwidth}
        \centering
        \includegraphics[width=\textwidth]{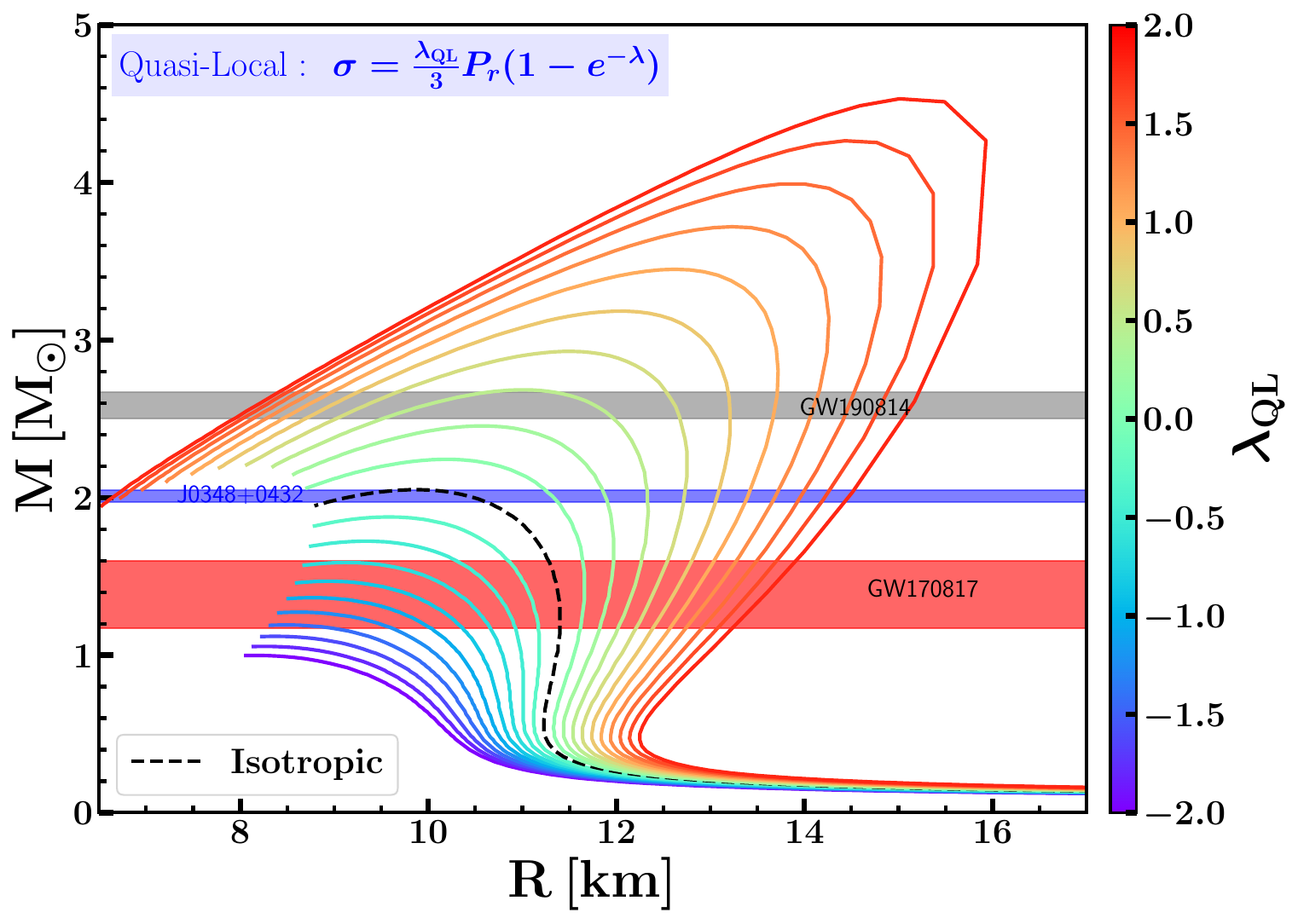}
        \caption{}
        \label{fig:QL_MR}
    \end{subfigure}
    \vskip\baselineskip
    \begin{subfigure}[t]{0.48\textwidth}
        \centering
        \includegraphics[width=\textwidth]{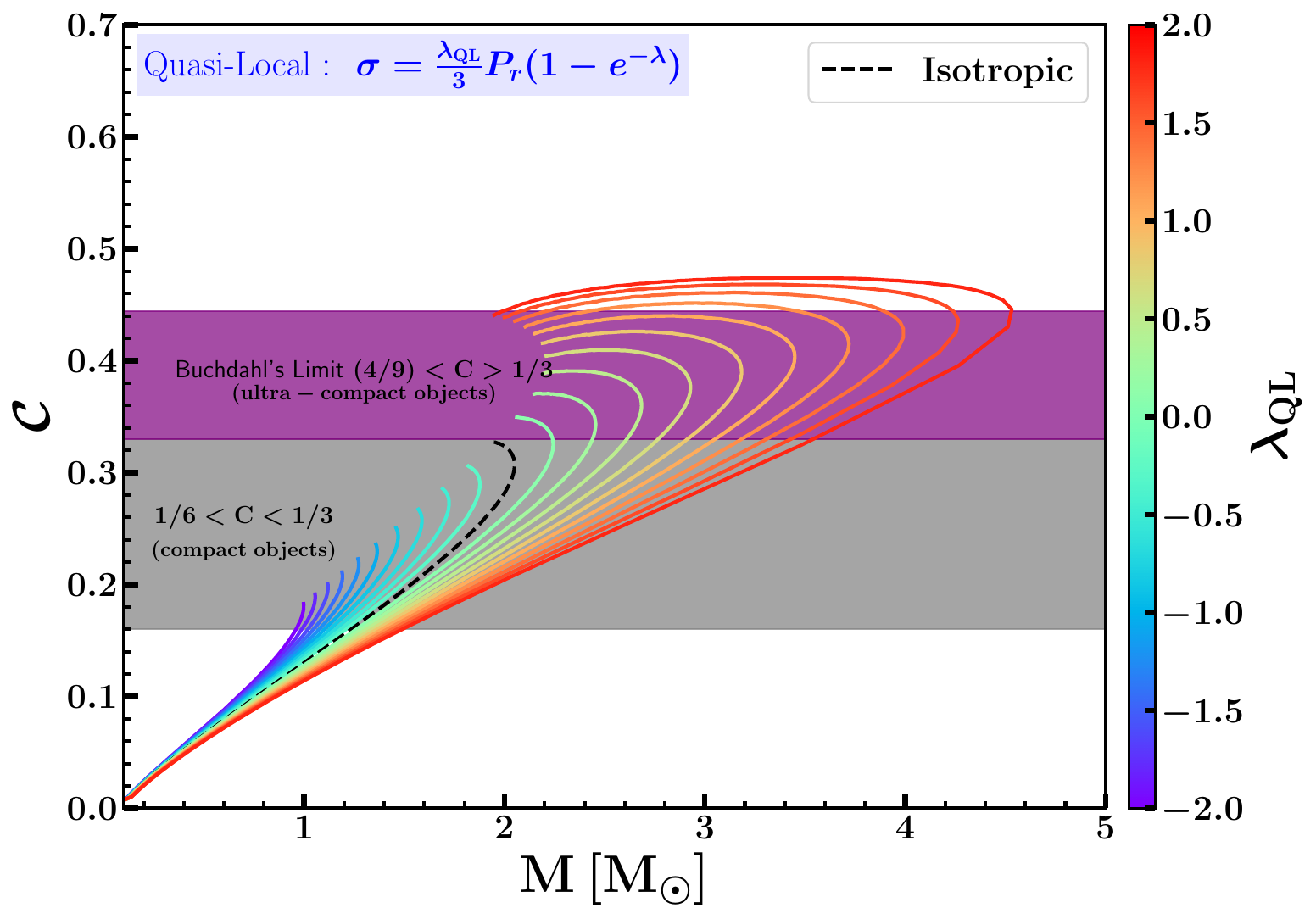}
        \caption{}
        \label{fig:QL_CM}
    \end{subfigure}
    \hfill
    \begin{subfigure}[t]{0.48\textwidth}
        \centering
        \includegraphics[width=\textwidth]{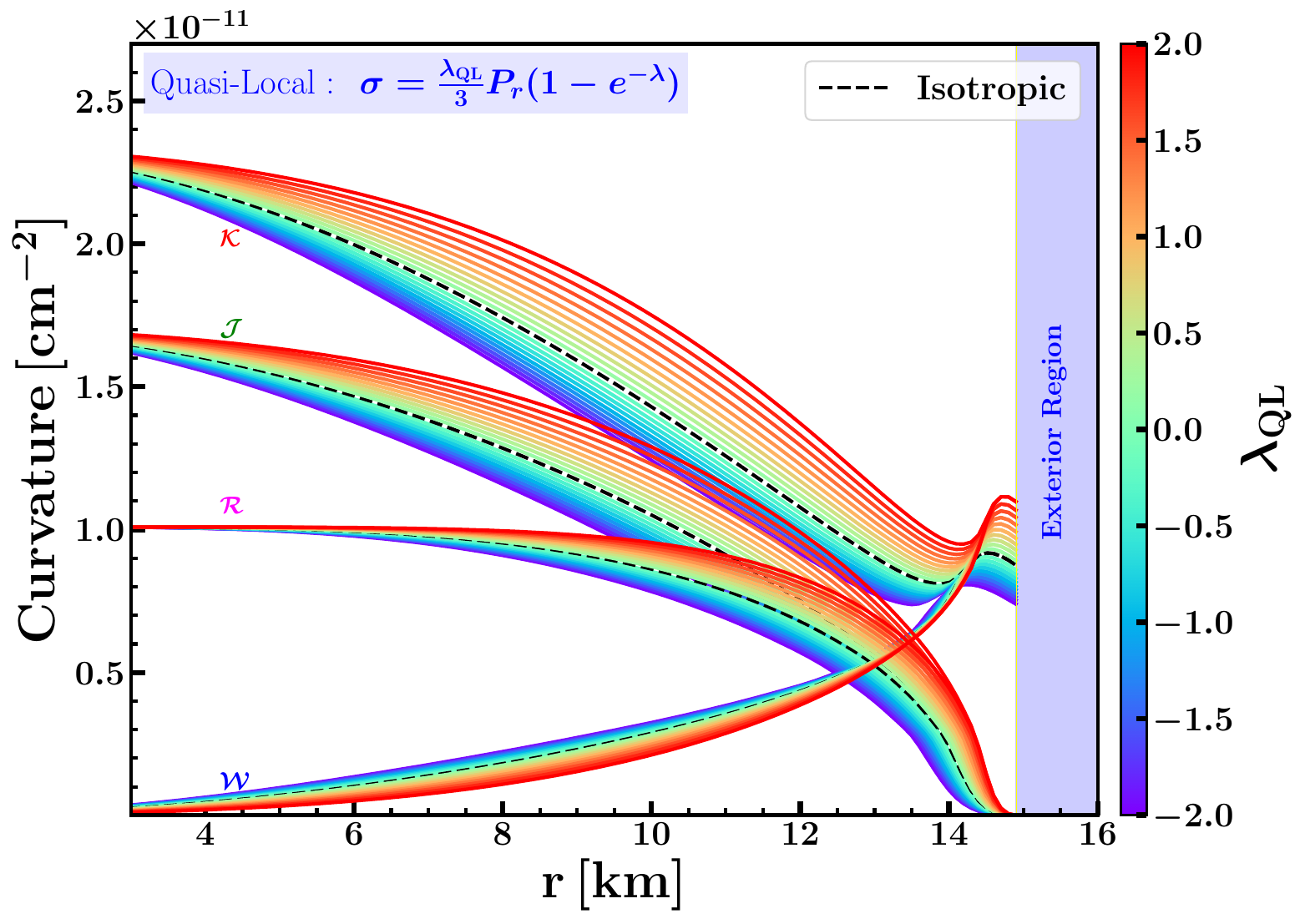}
        \caption{}
        \label{fig:QL_IC}
    \end{subfigure}
     \caption{\textbf{Upper left (a):} Anisotropy profile in the Quasi-Local model. 
     \textbf{Upper right (b):} relations between mass and radius.  The observed mass constraints from NICER data and GW observations are shown by the horizontal bands. Compactness ($\mathcal{C}= M/R$) as a function of mass is shown in \textbf{Lower left (c):}. The upper limit from the Buchdahl limit ($M/R \leq 4/9$) \cite{buchdahl_1959} is shown by the purple band. 
     \textbf{Lower right (d):} Curvature profile for various values of the parameter $\lambda_{\mathrm{QL}}$ as a function of radial coordinate.}
      \label{fig:QL_all}
\end{figure}

The QL anisotropy model proposed by Horvat et al.~\cite{horvat_2010} is briefly reviewed in this section. With this method, the anisotropy is expressed as a function of local stellar variables and can be expressed as
 \beq 
   \sigma = P_{\perp}-P_{r} = \frac{\lambda_{\mathrm{QL}}}{3}P_r \mu = \frac{\lambda_{\mathrm{QL}}}{3} P_r (1 - e^{-\lambda}) \, ,   
\eeq
 where the QL variable $\mu = 1 - e^{-\lambda} = \frac{2m(r)}{r}$ indicates the local compactness, and $\lambda_{\mathrm{QL}}$ measures the degree of anisotropy in the fluid. Using a similar range of the anisotropy parameter employed for the BL model, i.e., $\lambda_{\mathrm{QL}} \in [{-2.0, +2.0}]$, we present the anisotropy factor ($\sigma$) as a function of radius, the mass–radius relation, the variation of compactness with mass, and the curvature quantities for the SLy EoS in Fig.~\ref{fig:QL_all}.

\acknowledgments
ACK gratefully acknowledges the Inter-University Center for Astronomy and Astrophysics (IUCAA), Pune, India, for the facilities and hospitality provided during the course of this work. 

%\paragraph{Note added.} This is also a good position for notes added after the paper has been written.

\section{Evaluation of $\mathcal{W}$ and $\sigma$ in the vicinity of the origin}\label{APD:W_sigma}

\subsection{$\mathcal{W}$ near the origin}

The Weyl tensor's full contraction,
\begin{equation}
{\cal W}^2 \equiv {\cal C}^{\mu \nu \rho \sigma }{\cal C}_{\mu \nu \rho \sigma } 
\end{equation}
as seen in Eq.~(\ref{eqW}), it goes to zero near the center of the star.  In practice, determining
\begin{equation}\label{W}
{\cal W} = \frac{2}{\sqrt{3}}  \left(\frac{6m}{ r^3} - \kappa \cE\right)
\end{equation}
at exceptionally small radii ($r < 1$ cm) require careful examination, as the two contributions within the brackets essentially neutralize each other.  This almost-cancellation causes subtraction errors that are larger than the actual dimension of ${\cal W}$.  To solve this problem, we used a series expansion of ${\cal W}$ around the origin.

Because the system is spherically symmetric, the physical quantities pressure $P(r)$ and energy density $\cE(r)$ are analytic even functions of the radial coordinate $r$.  The mass function $m(r)$, defined as the integral of the energy density within a radius $r$, is an analytic odd function.  So, we use the following power series expansions around the origin: 
\begin{align}
P(r) &= P_c + \frac{1}{2}P''(0) r^2 + \frac{1}{24}P^{(4)}(0) r^4 +  \mathcal{O}(r^6), \label{eq:P_exp} \\[6pt]
\cE(r) &= \cE_c + \frac{1}{2}\cE''(0) r^2 + \frac{1}{24}\cE^{(4)}(0) r^4 +  \mathcal{O}(r^6), \label{eq:E_exp} \\[6pt]
m(r) &= \frac{1}{6}m'''(0) r^3 + \frac{1}{120}m^{(5)}(0) r^5 +  \mathcal{O}(r^7). \label{eq:m_exp}
\end{align}
The boundary conditions are carried out as follows: 
\[ P(0) = P_{\mathrm{c}}, \quad \mathcal{E}(0) = \mathcal{E}_{\mathrm{c}}, \quad m(0) = 0.\]

The first relevant TOV equation is given by:
\begin{equation}
m'(r) = 4\pi r^2 \cE(r). \label{eq:mass_TOV}
\end{equation}
Differentiating the expansion for $m(r)$ from Eq.~\eqref{eq:m_exp} yields:
\begin{equation} 
m'(r) = \frac{1}{2}m'''(0) r^2 + \frac{1}{24}m^{(5)}(0) r^4 +  \mathcal{O}(r^6).
\end{equation}
Substituting the expansion for $\cE(r)$ from Eq.~\eqref{eq:E_exp} into the right-hand side of Eq.~\eqref{eq:mass_TOV} gives:
\begin{equation}
4\pi r^2 \cE(r) = 4\pi r^2 \left( \cE_c + \frac{1}{2}\cE''(0) r^2 +  \mathcal{O}(r^4) \right) = 4\pi \cE_c r^2 + 2\pi \cE''(0) r^4 +  \mathcal{O}(r^6).
\end{equation}
Equating the coefficients for like powers of $r$ from both expressions for $m'(r)$:
\begin{itemize}
\item Coefficient of $r^2$:
  \begin{equation}
  \frac{1}{2}m'''(0) = 4\pi \cE_c \quad \Rightarrow \quad m'''(0) = 8\pi \cE_c,
  \end{equation}
\item Coefficient of $r^4$:
  \begin{equation}
  \frac{1}{24}m^{(5)}(0) = 2\pi \cE''(0) \quad \Rightarrow \quad m^{(5)}(0) = 48\pi \cE''(0).
  \end{equation}
\end{itemize}
Thus, the mass function expansion \eqref{eq:m_exp} becomes:
\begin{equation}
m(r) = \frac{4\pi}{3} \cE_c r^3 + \frac{2\pi}{5} \cE''(0) r^5 + \mathcal{O}(r^7).
 \label{eq:m_final}
\end{equation}

The hydrostatic equilibrium is governed by:
\begin{equation}
P'(r) = - \frac{m(r) \cE(r)}{r^2} \xi(r), \label{eq:TOV_P}
\end{equation}
where the relativistic factor $\xi(r)$ is defined as
\begin{equation}
\xi(r) = \left( 1 + \frac{P(r)}{\cE(r)} \right) \left( 1 + \frac{4\pi r^3 P(r)}{m(r)} \right) \left( 1 - \frac{2m(r)}{r} \right)^{-1}.
\label{eq:xi_def}
\end{equation}
Unlike what is sometimes assumed in approximate treatments, the central value of $\xi(r)$ is not unity. Evaluating Eq.~\eqref{eq:xi_def} at the center requires careful treatment of the factor containing $m(r)$ in the denominator. Using the leading-order behavior $m(r) = (4\pi/3)\cE_c r^3 + \mathcal{O}(r^5)$, we find
\begin{equation}
\frac{4\pi r^3 P(r)}{m(r)} = \frac{4\pi r^3 (P_c + \mathcal{O}(r^2))}{\frac{4\pi}{3}\cE_c r^3 + \mathcal{O}(r^5)} = \frac{3P_c}{\cE_c} + \mathcal{O}(r^2).
\end{equation}
The third factor approaches unity since $2m(r)/r = (8\pi/3)\cE_c r^2 + \mathcal{O}(r^4) \to 0$. Consequently, the central value is
\begin{equation}
\xi(0) = \xi_c \equiv \left(1 + \frac{P_c}{\cE_c}\right)\left(1 + \frac{3P_c}{\cE_c}\right) \neq 1.
\label{eq:xi_c}
\end{equation}
As an even function, $\xi(r)$ admits the expansion
\begin{equation}
\xi(r) = \xi_c + \frac{1}{2}\xi''(0) r^2 + \mathcal{O}(r^4) = \xi_c + \mathcal{O}(r^2).
\label{eq:xi_exp}
\end{equation}

Differentiating the pressure expansion \eqref{eq:P_exp} gives:
\begin{equation}
P'(r) = P''(0) r + \frac{1}{6}P^{(4)}(0) r^3 + \mathcal{O}(r^5).
\end{equation}
To compute the right-hand side of Eq.~\eqref{eq:TOV_P} near the origin, we use the expansions \eqref{eq:m_final}, \eqref{eq:E_exp}, and \eqref{eq:xi_exp}:
\begin{align}
m(r) \cE(r) \xi(r) &= \left( \frac{4\pi}{3} \cE_c r^3 + \mathcal{O}(r^5) \right) \left( \cE_c + \mathcal{O}(r^2) \right) \left( \xi_c + \mathcal{O}(r^2) \right) \nonumber \\
&= \frac{4\pi}{3} \cE_c^2 \xi_c r^3 + \mathcal{O}(r^5).
\end{align}
Dividing by $r^2$:
\begin{equation}
\frac{m(r) \cE(r) \xi(r)}{r^2} = \frac{4\pi}{3} \cE_c^2 \xi_c r + \mathcal{O}(r^3),
\end{equation}
so that
\begin{equation}
- \frac{m(r) \cE(r) \xi(r)}{r^2} = -\frac{4\pi}{3} \cE_c^2 \xi_c r + \mathcal{O}(r^3).
\end{equation}
Equating this with the expansion for $P'(r)$:
\begin{equation}
P''(0) r + \frac{1}{6}P^{(4)}(0) r^3 + \mathcal{O}(r^5) = -\frac{4\pi}{3} \cE_c^2 \xi_c r + \mathcal{O}(r^3),
\end{equation}
and comparing the coefficient of $r$, we find:
\begin{equation}
P''(0) = -\frac{4\pi}{3} \cE_c^2 \xi_c.
\label{eq:P2_final}
\end{equation}
Therefore, the pressure expansion is:
\begin{equation}
P(r) = P_c - \frac{2\pi}{3} \cE_c^2 \xi_c r^2 + \mathcal{O}(r^4).
\label{eq:P_final}
\end{equation}

To find $\cE''(0)$, we use the relation provided by the equation of state, $P' = (\dd P/\dd \cE) \cE'$. From Eq.~\eqref{eq:P_final}, $P'(0) = 0$, which immediately implies $\cE'(0) = 0$. Differentiating the relation and evaluating at $r=0$ yields:
\begin{equation}
P''(0) = \left( \frac{\dd P}{\dd \cE} \right)_c \cE''(0),
\end{equation}
since $(\cE'(0))^2 = 0$. Defining the adiabatic index at the center as $\gamma_c = (\cE_c / P_c)(\dd P/\dd \cE)_c$, we have $(\dd P/\dd \cE)_c = \gamma_c (P_c / \cE_c)$. Substituting $P''(0)$ from Eq.~\eqref{eq:P2_final}:
\begin{equation}
-\frac{4\pi}{3} \cE_c^2 \xi_c = \gamma_c \frac{P_c}{\cE_c} \cE''(0),
\end{equation}
which gives:
\begin{equation}
\cE''(0) = -\frac{4\pi}{3} \frac{\cE_c^3}{\gamma_c P_c} \xi_c.
\label{eq:E2_final}
\end{equation}
Thus, the energy density expansion is:
\begin{equation}
\cE(r) = \cE_c - \frac{2\pi}{3} \frac{\cE_c^3}{\gamma_c P_c} \xi_c r^2 + \mathcal{O}(r^4).
\label{eq:E_final}
\end{equation}

Using the mass expansion \eqref{eq:m_final}:
\begin{equation}
\frac{m(r)}{r^3} = \frac{4\pi}{3} \cE_c + \frac{2\pi}{5} \cE''(0) r^2 + \mathcal{O} (r^4),
\end{equation}
\begin{equation}
\frac{6m}{r^3} = 8\pi \cE_c + \frac{12\pi}{5} \cE''(0) r^2 + \mathcal{O} (r^4).
\end{equation}
Using the energy density expansion \eqref{eq:E_final}:
\begin{equation}
8\pi \cE = 8\pi \cE_c + 4\pi \cE''(0) r^2 + \mathcal{O}(r^4).
\end{equation}
Subtracting the two expressions within the parentheses of Eq.~\eqref{W}:
\begin{align}
\frac{6m}{r^3} - 8\pi \cE &= \left( 8\pi \cE_c - 8\pi \cE_c \right) + \left( \frac{12\pi}{5} - 4\pi \right) \cE''(0) r^2 + \mathcal{O}(r^4) \nonumber \\
&= -\frac{8\pi}{5} \cE''(0) r^2 + \mathcal{O}(r^4).
\end{align}
Substituting this result into Eq.~\eqref{W}:
\begin{equation}
\cW(r) = \frac{2}{\sqrt{3}} \left( -\frac{8\pi}{5} \cE''(0) r^2 \right) + \mathcal{O} (r^4) = -\frac{16\pi}{5\sqrt{3}} \cE''(0) r^2 + \mathcal{O} (r^4).
\end{equation}
Finally, using the expression for $\cE''(0)$ from Eq.~\eqref{eq:E2_final}, we obtain the leading-order behavior of the Weyl scalar near the origin:
\begin{equation}
\boxed{
\cW(r\rightarrow 0) = \frac{64\pi^2}{15\sqrt{3}} \frac{\cE_c^3}{\gamma_c P_c} \left(1 + \frac{P_c}{\cE_c}\right)\left(1 + \frac{3P_c}{\cE_c}\right) r^2.
} \label{eq:W_final}
\end{equation}
The adiabatic index at the central density for the SLy EoS is $\gamma_{\rm c} = (\mathcal{E}_{\rm c}/P_{\rm c})\,\Delta P/\Delta \mathcal{E}$, which gives $\gamma_{\rm c} \simeq 2.61$. 

\subsection{$\sigma$ near the origin}

The tangential pressure equation (refer to Eq.~(\ref{Anisotropy_eos})) is restated as follows:
\begin{equation}
    P_\bot = P_r + \frac{\lambda_{\rm BL}}{3} \frac{(\mathcal{E} + 3P_r)(\mathcal{E} + P_r) r^2}{1 - \frac{2m}{r}}\,,
    \label{eq:sigma_original}
\end{equation}
Using Eq.~(\ref{eq:sigma_original}), we can define the anisotropy function as
\begin{equation}
    \sigma(r) = P_\bot(r) - P_r(r) = \frac{\lambda_{\rm BL}}{3} \frac{(\mathcal{E} + 3P_r)(\mathcal{E} + P_r) r^2}{1 - \frac{2m}{r}}\,.
\end{equation}
We extend the appropriate parameters in the Taylor series in order to examine the behavior closer to the stellar center $(r \to 0)$:
\begin{align}
    \mathcal{E}(r) &= \mathcal{E}_c + \frac{1}{2} \mathcal{E}''(0) r^2 + \mathcal{O}(r^4), \\
    P_r(r) &= P_c - \frac{2\pi}{3} \mathcal{E}_c^2 \xi_c r^2 + \mathcal{O}(r^4), \\
    m(r) &= \frac{4\pi}{3} \mathcal{E}_c r^3 + \frac{2\pi}{5} \mathcal{E}''(0) r^5 + \mathcal{O}(r^7),
\end{align}
where $\mathcal{E}_c$ and $P_c$ are the central energy-density and pressure, respectively, and $\xi_c$ is precisely the same central factor that appeared in the $\mathcal{W}$ derivation:
\begin{equation}
    \xi_c = \left(1 + \frac{P_c}{\mathcal{E}_c}\right)\left(1 + \frac{3P_c}{\mathcal{E}_c}\right).
\end{equation}
Note the consistency: the same $\xi_c$ appears both in the pressure expansion above and in the Weyl scalar expression \eqref{eq:W_final}.

Expand the numerator and denominator of Eq.~\eqref{eq:sigma_original} individually.  The numerator becomes
\begin{equation}
    (\mathcal{E} + 3P_r)(\mathcal{E} + P_r) r^2 = \alpha r^2 + \mathcal{O}(r^4),
\end{equation}
with the constant coefficient
\begin{equation}
    \alpha = (\mathcal{E}_c + 3P_c)(\mathcal{E}_c + P_c).
\end{equation}
The denominator expands as
\begin{equation}
    \left(1 - \frac{2m}{r} \right)^{-1} = 1 + \frac{8\pi}{3} \mathcal{E}_c r^2 + \mathcal{O}(r^4).
\end{equation}
The leading-order behavior of the anisotropy function is obtained by combining the numerator and denominator:
\begin{align}
    \sigma(r) &= \frac{\lambda_{\rm BL}}{3} \left( \alpha r^2 + \alpha \cdot \frac{8\pi}{3} \mathcal{E}_c r^4 + \mathcal{O}(r^6) \right), \\
    &= \sigma_2 \cdot r^2 + \sigma_4 \cdot r^4 + \mathcal{O}(r^6),
\end{align}
where
\begin{align}
    \sigma_2 &= \frac{\lambda_{\rm BL}}{3} (\mathcal{E}_c + 3P_c)(\mathcal{E}_c + P_c), \\
    \sigma_4 &= \frac{8\pi \mathcal{E}_c}{3} \sigma_2.
\end{align}
Thus, the anisotropy function is quadratic near the center.
\begin{equation}
    \boxed{
    \sigma(r \to 0) = \sigma_2 \cdot r^2 + \sigma_4 \cdot r^4 + \mathcal{O}(r^6)}.
\end{equation}

\bibliographystyle{JHEP}
\bibliography{ref.bib}

@article{lattimer_2004,
  title={The physics of neutron stars},
  author={Lattimer, James M and Prakash, Maddappa},
  eprint ="astro-ph/0405262",
  doi ="10.1126/science.1090720",
  journal={science},
  volume={304},
  number={5670},
  pages={536--542},
  year={2004},
  publisher={American Association for the Advancement of Science}
}

@article{potekhin_2010,
  title={The physics of neutron stars},
  author={Potekhin, Aleksandr Y},
  eprint ="1102.5735 ",
  doi = "10.3367/UFNe.0180.201012c.1279",
  journal={Physics-Uspekhi},
  volume={53},
  number={12},
  pages={1235},
  year={2010},
  publisher={IOP Publishing}
}

@article{herrera_1997,
  title={Local anisotropy in self-gravitating systems},
  author={Herrera, Luis and Santos, Nilton O},
  doi = "10.1016/S0370-1573(96)00042-7",
  journal={Physics Reports},
  volume={286},
  number={2},
  pages={53--130},
  year={1997},
  publisher={Elsevier}
}

@article{yazadjiev_2012,
  title={Relativistic models of magnetars: Nonperturbative analytical approach},
  author={Yazadjiev, Stoytcho S},
 eprint ="1111.3536",
 doi ="10.1103/PhysRevD.85.044030",
  journal={Physical Review D—Particles, Fields, Gravitation, and Cosmology},
  volume={85},
  number={4},
  pages={044030},
  year={2012},
  publisher={APS}
}

@article{folomeev_2015,
  title={Magnetic fields in anisotropic relativistic stars},
  author={Folomeev, Vladimir and Dzhunushaliev, Vladimir},
 eprint= "1501.06275",
 doi ="10.1103/PhysRevD.91.044040",
  journal={Physical Review D},
  volume={91},
  number={4},
  pages={044040},
  year={2015},
  publisher={APS}
}

@article{kamiab_2015,
  title={The mass and radii of strongly magnetized neutron stars},
  author={Kamiab, Farbod and Broderick, Avery E and Afshordi, Niayesh},
 eprint="1503.03898",
  journal={arXiv preprint arXiv:1503.03898},
  year={2015}
}

@article{ioka_2004,
  title={Relativistic stars with poloidal and toroidal magnetic fields and meridional flow},
  author={Ioka, Kunihito and Sasaki, Misao},
 eprint ="astro-ph/0305352",
doi = "10.1086/379650",
  journal={The Astrophysical Journal},
  volume={600},
  number={1},
  pages={296},
  year={2004},
  publisher={IOP Publishing}
}

@article{frieben_2012,
  title={Equilibrium models of relativistic stars with a toroidal magnetic field},
  author={Frieben, Joachim and Rezzolla, Luciano},
 eprint ="1207.4035",
 doi ="10.1111/j.1365-2966.2012.22027.x",
  journal={Monthly Notices of the Royal Astronomical Society},
  volume={427},
  number={4},
  pages={3406--3426},
  year={2012},
  publisher={Blackwell Science Ltd Oxford, UK}
}

@article{carter_1998,
  title={Relativistic models for superconducting-superfluid mixtures},
  author={Carter, Brandon and Langlois, David},
  eprint ="gr-qc/9806024",
  doi="10.1016/S0550-3213(98)00430-1",
  journal={Nuclear Physics B},
  volume={531},
  number={1-3},
  pages={478--504},
  year={1998},
  publisher={Elsevier}
}

@article{lombardo_2001,
  title={Superfluidity in neutron star matter},
  author={Lombardo, Umberto and Schulze, Hans-Josef},
  eprint="astro-ph/0012209",
  doi ="10.1007/3-540-44578-1_2",
  journal={Physics of Neutron Star Interiors},
  pages={30--53},
  year={2001},
  publisher={Springer}
}

@article{heiselberg_2000,
  title={Phases of dense matter in neutron stars},
  author={Heiselberg, Henning and Hjorth-Jensen, Morten},
  eprint ="nucl-th/9902033",
  doi = "10.1016/S0370-1573(99)00110-6",
  journal={Physics Reports},
  volume={328},
  number={5-6},
  pages={237--327},
  year={2000},
  publisher={Elsevier}
}

@article{baym_1969,
  title={Superfluidity in neutron stars},
  author={Baym, Gordon and Pethick, Christopher and Pines, David},
  doi = "10.1038/224673a0",
  journal={Nature},
  volume={224},
  number={5220},
  pages={673--674},
  year={1969},
  publisher={Nature Publishing Group UK London}
}

@article{haskell_2018,
  title={Superfluidity and superconductivity in neutron stars},
  author={Haskell, Brynmor and Sedrakian, Armen},
 eprint = "	1709.10340 ",
 doi ="10.1007/978-3-319-97616-7_8",
  journal={The Physics and Astrophysics of Neutron Stars},
  pages={401--454},
  year={2018},
  publisher={Springer}
}

@article{pines_1985,
  title={Superfluidity in neutron stars},
  author={Pines, David and Alpar, M Ali},
  doi = "10.1038/316027a0",
  journal={Nature},
  volume={316},
  number={6023},
  pages={27--32},
  year={1985},
  publisher={Nature Publishing Group UK London}
}

@article{nelmes_2012,
  title={Phase transition and anisotropic deformations of neutron star matter},
  author={Nelmes, Susan and Piette, Bernard MAG},
 eprint = "	1204.0910 ",
 doi = "10.1103/PhysRevD.85.123004",
journal={Physical Review D—Particles, Fields, Gravitation, and Cosmology},
  volume={85},
  number={12},
  pages={123004},
  year={2012},
  publisher={APS}
}

@article{canuto_1974,
  title={Crystallization of dense neutron matter},
  author={Canuto, V and Chitre, SM},
  doi ="10.1103/PhysRevD.9.1587",
  journal={Physical Review D},
  volume={9},
  number={6},
  pages={1587},
  year={1974},
  publisher={APS}
}

@article{cromartie_2020,
  title={Relativistic Shapiro delay measurements of an extremely massive millisecond pulsar},
  author={Cromartie, H Thankful and Fonseca, Emmanuel and Ransom, Scott M and Demorest, Paul B and Arzoumanian, Zaven and Blumer, Harsha and Brook, Paul R and DeCesar, Megan E and Dolch, Timothy and Ellis, Justin A and others},
 eprint ="	1904.06759",
 doi ="https://doi.org/10.1038/s41550-019-0880-2",
  journal={Nature Astronomy},
  volume={4},
  number={1},
  pages={72--76},
  year={2020},
  publisher={Nature Publishing Group UK London}
}

@article{fonseca_2021,
  title={Refined mass and geometric measurements of the high-mass PSR J0740+ 6620},
  author={Fonseca, Emmanuel and Cromartie, H Thankful and Pennucci, Timothy T and Ray, Paul S and Kirichenko, A Yu and Ransom, Scott M and Demorest, Paul B and Stairs, Ingrid H and Arzoumanian, Zaven and Guillemot, Lucas and others},
 eprint = "2104.00880",
 doi = "10.3847/2041-8213/ac03b8",
  journal={The Astrophysical Journal Letters},
  volume={915},
  number={1},
  pages={L12},
  year={2021},
  publisher={IOP Publishing}
}

@article{li_2019,
  title={Towards understanding astrophysical effects of nuclear symmetry energy},
  author={Li, Bao-An and Krastev, Plamen G and Wen, De-Hua and Zhang, Nai-Bo},
 eprint = "1905.13175",
 doi = "10.1140/epja/i2019-12780-8",
 journal={The European Physical Journal A},
  volume={55},
  number={7},
  pages={117},
  year={2019},
  publisher={Springer}
}

@article{lim_2019,
  title={Bayesian modeling of the nuclear equation of state for neutron star tidal deformabilities and GW170817},
  author={Lim, Y and Holt, JW},
  eprint = "1902.05502",
 doi = "10.1140/epja/i2019-12917-9",
  journal={The European Physical Journal A},
  volume={55},
  number={11},
  pages={209},
  year={2019},
  publisher={Springer}
}

@article{malik_2018,
  title={GW170817: Constraining the nuclear matter equation of state from the neutron star tidal deformability},
  author={Malik, Tuhin and Alam, N and Fortin, M and Provid{\^e}ncia, C and Agrawal, BK and Jha, TK and Kumar, Bharat and Patra, SK},
  eprint ="1805.11963",
  doi = "10.1103/PhysRevC.98.035804",
  journal={Physical Review C},
  volume={98},
  number={3},
  pages={035804},
  year={2018},
  publisher={APS}
}

@article{ozel_2016,
  title={Masses, radii, and the equation of state of neutron stars},
  author={{\"O}zel, Feryal and Freire, Paulo},
  eprint ="1603.02698",
  doi = "10.1146/annurev-astro-081915-023322",
  journal={Annual Review of Astronomy and Astrophysics},
  volume={54},
  number={1},
  pages={401--440},
  year={2016},
  publisher={Annual Reviews}
}

@article{brandes_2025,
  title={Implications of latest NICER data for the neutron star equation of state},
  author={Brandes, Len and Weise, Wolfram},
  eprint ="	2412.05923",
  doi = "10.1103/PhysRevD.111.034005",
  journal={Physical Review D},
  volume={111},
  number={3},
  pages={034005},
  year={2025},
  publisher={APS}
}

@article{abbott_2017,
  title={GW170817: observation of gravitational waves from a binary neutron star inspiral},
  author={Abbott, Benjamin P and Abbott, Rich and Abbott, Thomas D and Acernese, Fausto and Ackley, Kendall and Adams, Carl and Adams, Thomas and Addesso, Paolo and Adhikari, Rana X and Adya, Vaishali B and others},
  eprint ="1710.05832 ",
  doi ="10.1103/PhysRevLett.119.161101",
  journal={Physical review letters},
  volume={119},
  number={16},
  pages={161101},
  year={2017},
  publisher={APS}
}

@article{abbott_2018,
  title={GW170817: Measurements of neutron star radii and equation of state},
  author={Abbott, Benjamin P and Abbott, Richard and Abbott, TD and Acernese, F and Ackley, K and Adams, C and Adams, T and Addesso, P and Adhikari, Rana X and Adya, Vaishali B and others},
  eprint ="1805.11581 ",
 doi = "10.1103/PhysRevLett.121.161101",
  journal={Physical review letters},
  volume={121},
  number={16},
  pages={161101},
  year={2018},
  publisher={APS}
}

@article{abbott_2019,
  title={Properties of the binary neutron star merger GW170817},
  author={Abbott, BPea and Abbott, R and Abbott, TD and Acernese, F and Ackley, K and Adams, C and Adams, T and Addesso, P and Adhikari, RX and Adya, VB and others},
  eprint ="1805.11579",
  doi = "10.1103/PhysRevX.9.011001",
  journal={Physical Review X},
  volume={9},
  number={1},
  pages={011001},
  year={2019},
  publisher={APS}
}

@article{abbott_2020,
  title={GW190814: gravitational waves from the coalescence of a 23 solar mass black hole with a 2.6 solar mass compact object},
  author={Abbott, Richard and Abbott, TD and Abraham, S and Acernese, Fausto and Ackley, K and Adams, C and Adhikari, Rana X and Adya, VB and Affeldt, Christoph and Agathos, Michail and others},
  eprint = "2006.12611",
  doi = "10.3847/2041-8213/ab960f",
  journal={The Astrophysical Journal Letters},
  volume={896},
  number={2},
  pages={L44},
  year={2020},
  publisher={IOP Publishing}
}

@article{hossain_2021a,
  title={Equation of states in the curved spacetime of spherical degenerate stars},
  author={Hossain, Golam Mortuza and Mandal, Susobhan},
eprint = "	2005.08783",
 doi = "10.1088/1475-7516/2021/02/026",
  journal={Journal of Cosmology and Astroparticle Physics},
  volume={2021},
  number={02},
  pages={026},
  year={2021},
  publisher={IOP Publishing}
}

@article{hossain_2021b,
  title={Higher mass limits of neutron stars from the equation of states in curved spacetime},
  author={Hossain, Golam Mortuza and Mandal, Susobhan},
  eprint = "	2109.09606 ",
  doi ="10.1103/PhysRevD.104.123005",
  journal={Physical Review D},
  volume={104},
  number={12},
  pages={123005},
  year={2021},
  publisher={APS}
}

@article{hossain_2022,
  title={Equation of states in the curved spacetime of slowly rotating degenerate stars},
  author={Hossain, Golam Mortuza and Mandal, Susobhan},
 eprint = "2204.12352",
doi = " 10.1088/1475-7516/2022/10/008",
  journal={Journal of Cosmology and Astroparticle Physics},
  volume={2022},
  number={10},
  pages={008},
  year={2022},
  publisher={IOP Publishing}
}

@article{hossain_2023,
  title={Effects of magnetic field on the equation of state in curved spacetime of a neutron star},
  author={Hossain, Golam Mortuza and Mandal, Susobhan},
  eprint = "	2312.16589",
  journal={arXiv preprint arXiv:2312.16589},
  year={2023}
}

@article{li_2022,
  title={Do we need dense matter equation of state in curved spacetime for neutron stars?},
  author={Li, Jianing and Guo, Tao and Zhao, Jiaxing and He, Lianyi},
  eprint = "2206.02106",
  doi = "10.1103/PhysRevD.106.083021",
  journal={Physical Review D},
  volume={106},
  number={8},
  pages={083021},
  year={2022},
  publisher={APS}
}

@article{eksi_2014,
  title={What does a measurement of mass and/or radius of a neutron star constrain: Equation of state or gravity?},
  author={Ek{\c{s}}i, Kaz{\i}m Yavuz and G{\"u}ng{\"o}r, Can and T{\"u}rko{\u{g}}lu, Murat Metehan},
  eprint ="1402.0488 ",
  doi = "10.1103/PhysRevD.89.063003",
  journal={Physical Review D},
  volume={89},
  number={6},
  pages={063003},
  year={2014},
  publisher={APS}
}

@article{walecka_1974,
  title={A theory of highly condensed matter},
  author={Walecka, John Dirk},
  doi = "10.1016/0003-4916(74)90208-5",
  journal={Annals of Physics},
  volume={83},
  number={2},
  pages={491--529},
  year={1974},
  publisher={Elsevier}
}

@article{bowers_1974,
  title={Anisotropic spheres in general relativity},
  author={Bowers, Richard L and Liang, EPT},
  doi ="10.1086/152760",
  journal={Astrophysical Journal, Vol. 188, p. 657 (1974)},
  volume={188},
  pages={657},
  year={1974}
}

@article{das_2022,
  title={I-Love-C relation for an anisotropic neutron star},
  author={Das, HC},
 eprint="2208.12566",
 doi = "10.1103/PhysRevD.106.103518",
 journal={Physical Review D},
  volume={106},
  number={10},
  pages={103518},
  year={2022},
  publisher={APS}
}

@article{das_2021,
  title={Impacts of dark matter on the curvature of the neutron star},
  author={Das, HC and Kumar, Ankit and Kumar, Bharat and Biswal, SK and Patra, SK},
  eprint ="2007.05382 ",
  doi = "10.1088/1475-7516/2021/01/007",
  journal={Journal of Cosmology and Astroparticle Physics},
  volume={2021},
  number={01},
  pages={007},
  year={2021},
  publisher={IOP Publishing}
}

@article{regge_1957,
  title={Stability of a Schwarzschild singularity},
  author={Regge, Tullio and Wheeler, John A},
  doi = "10.1103/PhysRev.108.1063",
  journal={Physical Review},
  volume={108},
  number={4},
  pages={1063},
  year={1957},
  publisher={APS}
}

@article{thorne_1967,
  title={Non-radial pulsation of general-relativistic stellar models. I. Analytic analysis for L>= 2},
  author={Thorne, Kip S and Campolattaro, Alfonso},
 doi ="10.1086/149586 ",
  journal={Astrophysical Journal, vol. 149, p. 591},
  volume={149},
  pages={591},
  year={1967}
}

@article{hinderer_2008,
  title={Tidal Love numbers of neutron stars},
  author={Hinderer, Tanja},
  eprint  = "0711.2420",
  doi = "10.1086/533487",
  journal={The Astrophysical Journal},
  volume={677},
  number={2},
  pages={1216},
  year={2008},
  publisher={IOP Publishing}
}

@article{biswas_2019,
  title={Tidal deformability of an anisotropic compact star: Implications of GW170817},
  author={Biswas, Bhaskar and Bose, Sukanta},
  eprint = "1903.04956",
  doi = "10.1103/PhysRevD.99.104002",
  journal={Physical Review D},
  volume={99},
  number={10},
  pages={104002},
  year={2019},
  publisher={APS}
}

@article{damour_2009,
  title={Relativistic tidal properties of neutron stars},
  author={Damour, Thibault and Nagar, Alessandro},
eprint="0906.0096",
doi = "10.1103/PhysRevD.80.084035",
  journal={Physical Review D—Particles, Fields, Gravitation, and Cosmology},
  volume={80},
  number={8},
  pages={084035},
  year={2009},
  publisher={APS}
}

@article{hinderer_2009,
  title={Erratum:" tidal love numbers of neutron stars"(2008, apj, 677, 1216)},
  author={Hinderer, Tanja},
  journal={Astrophysical Journal},
  volume={697},
  number={1},
  pages={964},
  year={2009},
  publisher={Citeseer}
}

@article{idrisy_2015,
  title={R-mode frequencies of slowly rotating relativistic neutron stars with realistic equations of state},
  author={Idrisy, Ashikuzzaman and Owen, Benjamin J and Jones, David I},
 eprint = "1410.7360",
 doi = "10.1103/PhysRevD.91.024001",
  journal={Physical Review D},
  volume={91},
  number={2},
  pages={024001},
  year={2015},
  publisher={APS}
}

@article{romani_2022,
  title={PSR J0952- 0607: The fastest and heaviest known galactic neutron star},
  author={Romani, Roger W and Kandel, D and Filippenko, Alexei V and Brink, Thomas G and Zheng, WeiKang},
  eprint="2207.05124",
  doi = "10.3847/2041-8213/ac8007",
  journal={The Astrophysical Journal Letters},
  volume={934},
  number={2},
  pages={L17},
  year={2022},
  publisher={IOP Publishing}
}

@article{riley_2019nicer,
  title={A NICER view of PSR J0030+ 0451: millisecond pulsar parameter estimation},
  author={Riley, Thomas E and Watts, Anna L and Bogdanov, Slavko and Ray, Paul S and Ludlam, Renee M and Guillot, Sebastien and Arzoumanian, Zaven and Baker, Charles L and Bilous, Anna V and Chakrabarty, Deepto and others},
  journal={The Astrophysical Journal Letters},
  eprint ="1912.05702",
  doi = "10.3847/2041-8213/ab481c",
  volume={887},
  number={1},
  pages={L21},
  year={2019},
  publisher={IOP Publishing}
}

@article{miller_2019psr,
  title={PSR J0030+ 0451 mass and radius from NICER data and implications for the properties of neutron star matter},
  author={Miller, MC and Lamb, Frederick K and Dittmann, AJ and Bogdanov, Slavko and Arzoumanian, Zaven and Gendreau, Keith C and Guillot, S and Harding, AK and Ho, WCG and Lattimer, JM and others},
eprint = "1912.05705",
doi = "10.3847/2041-8213/ab50c5",
  journal={The Astrophysical Journal Letters},
  volume={887},
  number={1},
  pages={L24},
  year={2019},
  publisher={IOP Publishing}
}

@article{miller_2021,
  title={The radius of PSR J0740+ 6620 from NICER and XMM-Newton data},
  author={Miller, M Coleman and Lamb, FK and Dittmann, AJ and Bogdanov, S and Arzoumanian, Z and Gendreau, KC and Guillot, S and Ho, WCG and Lattimer, JM and Loewenstein, M and others},
  eprint = "2105.06979",
  doi = "10.3847/2041-8213/ac089b",
  journal={The Astrophysical Journal Letters},
  volume={918},
  number={2},
  pages={L28},
  year={2021},
  publisher={IOP Publishing}
}

@article{abbott_2021,
  title={Observation of gravitational waves from two neutron star--black hole coalescences},
  author={Abbott, R and Abbott, Thomas D and Abraham, S and Acernese, Fausto and Ackley, K and Adams, A and Adams, C and Adhikari, RX and Adya, VB and Affeldt, Christoph and others},
eprint = "2106.15163 ",
  doi = "10.3847/2041-8213/ac082e",
  journal={The Astrophysical journal letters},
  volume={915},
  number={1},
  pages={L5},
  year={2021},
  publisher={IoP Publishing}
}

@article{antoniadis_2013,
  title={A massive pulsar in a compact relativistic binary},
  author={Antoniadis, John and Freire, Paulo CC and Wex, Norbert and Tauris, Thomas M and Lynch, Ryan S and Van Kerkwijk, Marten H and Kramer, Michael and Bassa, Cees and Dhillon, Vik S and Driebe, Thomas and others},
eprint= "1304.6875",
doi = "10.1126/science.1233232",
  journal={Science},
  volume={340},
  number={6131},
  pages={1233232},
  year={2013},
  publisher={American Association for the Advancement of Science}
}

@article{kumar_2019,
  title={Inferring neutron star properties from GW170817 with universal relations},
  author={Kumar, Bharat and Landry, Philippe},
 eprint = "1902.04557",
 doi = "10.1103/PhysRevD.99.123026",
  journal={Physical Review D},
  volume={99},
  number={12},
  pages={123026},
  year={2019},
  publisher={APS}
}

@article{landry_2020,
  title={Nonparametric constraints on neutron star matter with existing and upcoming gravitational wave and pulsar observations},
  author={Landry, Philippe and Essick, Reed and Chatziioannou, Katerina},
  eprint = "2003.04880",
  doi ="10.1103/PhysRevD.101.123007",
  journal={Physical Review D},
  volume={101},
  number={12},
  pages={123007},
  year={2020},
  publisher={APS}
}

@article{jiang_2020,
  title={PSR J0030+ 0451, GW170817, and the nuclear data: joint constraints on equation of state and bulk properties of neutron stars},
  author={Jiang, Jin-Liang and Tang, Shao-Peng and Wang, Yuan-Zhu and Fan, Yi-Zhong and Wei, Da-Ming},
 eprint = "1912.07467",
 doi ="10.3847/1538-4357/ab77cf",
  journal={The Astrophysical Journal},
  volume={892},
  number={1},
  pages={55},
  year={2020},
  publisher={IOP Publishing}
}

@article{abbott_2020gw190425,
 title={GW190425: Observation of a compact binary coalescence with total mass $\sim 3.4\,M_{\odot}$},
  author={Abbott, Benjamin P and Abbott, Robert and Abbott, TD and Abraham, S and Acernese, Fausto and Ackley, K and Adams, C and Adhikari, RX and Adya, VB and Affeldt, Christoph and others},
erpint ="2001.01761",
  doi = "10.3847/2041-8213/ab75f5",
  journal={The Astrophysical Journal},
  volume={892},
  number={1},
  pages={L3},
  year={2020},
  publisher={American Astronomical Society}
}

@article{flanagan_2008,
  title={Constraining neutron-star tidal Love numbers with gravitational-wave detectors},
  author={Flanagan, Eanna E and Hinderer, Tanja},
  doi ="10.1103/PhysRevD.77.021502",
  eprint = "0709.1915",
  journal={Physical Review D—Particles, Fields, Gravitation, and Cosmology},
  volume={77},
  number={2},
  pages={021502},
  year={2008},
  publisher={APS}
}

@article{psaltis_2008,
  title={Probes and tests of strong-field gravity with observations in the electromagnetic spectrum},
  author={Psaltis, Dimitrios},
  eprint= "0806.1531",
  doi = "10.12942/lrr-2008-9",
  journal={Living Reviews in Relativity},
  volume={11},
  number={1},
  pages={9},
  year={2008},
  publisher={Springer}
}

@article{rahmansyah_2020,
  title={Anisotropic neutron stars with hyperons: implication of the recent nuclear matter data and observations of neutron stars},
  author={Rahmansyah, A and Sulaksono, A and Wahidin, AB and Setiawan, AM},
  doi = "10.1140/epjc/s10052-020-8361-4",
  journal={The European Physical Journal C},
  volume={80},
  number={8},
  pages={769},
  year={2020},
  publisher={Springer}
}

@article{horvat_2010,
  title={Radial pulsations and stability of anisotropic stars with a quasi-local equation of state},
  author={Horvat, Dubravko and Iliji{\'c}, Sa{\v{s}}a and Marunovi{\'c}, Anja},
  eprint = "1010.0878",
  doi = "10.1088/0264-9381/28/2/025009",
  journal={Classical and quantum gravity},
  volume={28},
  number={2},
  pages={025009},
  year={2010},
  publisher={IOP Publishing}
}

@article{cosenza_1981,
  title={Some models of anisotropic spheres in general relativity},
  author={Cosenza, M and Herrera, L and Esculpi, M and Witten, L},
  doi = "10.1063/1.524742",
  journal={Journal of Mathematical Physics},
  volume={22},
  number={1},
  pages={118--125},
  year={1981},
  publisher={American Institute of Physics}
}

@article{becerra_2024,
  title={Realistic anisotropic neutron stars: Pressure effects},
  author={Becerra, LM and Becerra-Vergara, EA and Lora-Clavijo, FD},
  eprint ="2401.10311 ",
  doi = "10.1103/PhysRevD.109.043025",
  journal={Physical Review D},
  volume={109},
  number={4},
  pages={043025},
  year={2024},
  publisher={APS}
}

@article{mohanty_2024,
  title={The impact of anisotropy on neutron star properties: insights from $I-f-C$ universal relations},
  author={Mohanty, Sailesh Ranjan and Ghosh, Sayantan and Routaray, Pinku and Das, HC and Kumar, Bharat},
eprint= "2305.15724 ",
doi = "10.1088/1475-7516/2024/03/054",
  journal={Journal of Cosmology and Astroparticle Physics},
  volume={2024},
  number={03},
  pages={054},
  year={2024},
  publisher={IOP Publishing}
}

@article{das_2023,
  title={Constraining the surface curvature of an anisotropic neutron star},
  author={Das, HC and Pattnaik, Jeet Amrit and Patra, SK},
  eprint = "2301.12673",
  doi = "10.1103/PhysRevD.107.083007",
  journal={Physical Review D},
  volume={107},
  number={8},
  pages={083007},
  year={2023},
  publisher={APS}
}

@article{pretel_2022,
  title={Moment of inertia of slowly rotating anisotropic neutron stars in f (R, T) gravity},
  author={Pretel, Juan MZ},
  eprint = "2301.02881",
  doi = "10.1142/S0217732322501887",
  journal={Modern Physics Letters A},
  volume={37},
  number={28},
  pages={2250188},
  year={2022},
  publisher={World Scientific}
}

@article{buchdahl_1959,
  title={General relativistic fluid spheres},
  author={Buchdahl, Hans A},
  doi = "10.1103/PhysRev.116.1027",
  journal={Physical Review},
  volume={116},
  number={4},
  pages={1027},
  year={1959},
  publisher={APS}
}

@article{douchin_2001,
  title={A unified equation of state of dense matter and neutron star structure},
  author={Douchin, F and Haensel, P},
eprint="astro-ph/0111092",
  doi = "10.1051/0004-6361:20011402",
  journal={Astronomy \& Astrophysics},
  volume={380},
  number={1},
  pages={151--167},
  year={2001},
  publisher={EDP Sciences}
}

@book{glendenning_2012,
  title={Compact stars: Nuclear physics, particle physics and general relativity},
  author={Glendenning, Norman K},
  year={2012},
  publisher={Springer Science \& Business Media}
}

@article{rezzolla_2025,
  title={On the maximum compactness of neutron stars},
  author={Rezzolla, Luciano and Ecker, Christian},
 eprint = "2510.12870",
  doi = "10.48550/arXiv.2510.12870",
  journal={arXiv prep rint arXiv:2510.12870},
  year={2025}
}

@article{becerra_2025,
  title={On the Stability of Anisotropic Neutron Stars},
  author={Becerra, LM and Becerra-Vergara, EA and Lora-Clavijo, FD and Rodriguez, JF},
  eprint ="2512.19825",
  journal={arXiv preprint arXiv:2512.19825},
  year={2025}
}

@article{pretel_2020,
  title={Equilibrium, radial stability and non-adiabatic gravitational collapse of anisotropic neutron stars},
  author={Pretel, Juan MZ},
 eprint = "2008.05331 ",
 doi ="10.1140/epjc/s10052-020-8301-3",
  journal={The European Physical Journal C},
  volume={80},
  number={8},
  pages={726},
  year={2020},
  publisher={Springer}
}

@article{mohanty_2024u,
  title={Unstable anisotropic neutron stars: Probing the limits of gravitational collapse},
  author={Mohanty, Sailesh Ranjan and Ghosh, Sayantan and Kumar, Bharat},
  eprint ="2304.02439 ",
  doi = "10.1103/PhysRevD.109.123039",
  journal={Physical Review D},
  volume={109},
  number={12},
  pages={123039},
  year={2024},
  publisher={APS}
}

@article{cherubini_2002,
  title={Second order scalar invariants of the Riemann tensor: applications to black hole spacetimes},
  author={Cherubini, Christian and Bini, Donato and Capozziello, Salvatore and Ruffini, Remo},
  eprint ="gr-qc/0302095",
  doi="10.1142/S0218271802002037",
  journal={International Journal of Modern Physics D},
  volume={11},
  number={06},
  pages={827--841},
  year={2002},
  publisher={World Scientific}
}

@article{ghosh_2026,
  title={Spacetime curvature as a probe of exotic core phases in neutron stars within modified gravity},
  author={Ghosh, Sayantan and Kumar, Bharat and Mahapatra, Subhash},
  eprint = "2508.08866 ",
  doi = "10.1103/rjzl-pcr4",
  journal={Physical Review D},
  volume={113},
  number={2},
  pages={024070},
  year={2026},
  publisher={APS}
}

\end{document}